%% file: amnesty-paper.tex
\documentclass[hidelinks,12pt]{article}

\usepackage{graphicx,amsfonts,psfrag,layout,subcaption,array,longtable,lscape,booktabs,dcolumn,natbib,amsmath,amssymb,amssymb,amsthm,setspace,epigraph,chronology,color, colortbl,caption,wasysym,threeparttable,adjustbox,authblk}
\usepackage[]{graphicx}\usepackage[]{color}
\usepackage[page]{appendix}
\usepackage{hyperref, url} 
\usepackage[section]{placeins}
\usepackage[linewidth=1pt]{mdframed}
\usepackage[margin=1in]{geometry} 

\title{Amnesty Policy and Elite Persistence in the Postbellum South: Evidence from a Regression Discontinuity Design}
\author[ ]{Jason Poulos\thanks{\emph{Address for correspondence:} 180 Longwood Avenue, Boston, MA 02115. \emph{E-mail:} poulos@hcp.med.harvard.edu. \emph{Acknowledgements:} I thank Cyrus Corman-Gill, Amita Chauhan, Desiree Moshayedi, and JaVonte Morris-Wilson for help with transcribing census data. The paper benefited from constructive comments by David Bateman and participants of the ``Slavery \& Its Legacies Symposium" hosted by the USC Bedrosian Center. This work was partially supported by the National Science Foundation Graduate Research Fellowship under Grant DGE-1106400 and the National Science Foundation under Grant DMS-1638521 to the Statistical and Applied Mathematical Sciences Institute. Data and code are available at \url{https://github.com/jvpoulos/amnesty}.}}
\affil[ ]{Department of Health Care Policy, Harvard Medical School}
\date{}
\usepackage{xr}
\usepackage[bottom]{footmisc}

\usepackage{footmisc}
\DefineFNsymbols{mySymbols}{{\ensuremath\dagger}{\ensuremath\ddagger}\S\P
   *{**}{\ensuremath{\dagger\dagger}}{\ensuremath{\ddagger\ddagger}}}
\setfnsymbol{mySymbols}

\usepackage{tabularx}
\newcolumntype{Y}{>{\raggedleft\arraybackslash}X}

\definecolor{Gray}{gray}{0.9}

\newcommand{\captionfonts}{\scriptsize}

\makeatletter  
\long\def\@makecaption#1#2{%
  \vskip\abovecaptionskip
  \sbox\@tempboxa{{\captionfonts #1: #2}}%
  \ifdim \wd\@tempboxa >\hsize
    {\captionfonts #1: #2\par}
  \else
    \hbox to\hsize{\hfil\box\@tempboxa\hfil}%
  \fi
  \vskip\belowcaptionskip}
 
\newtheorem*{assumption*}{\assumptionnumber}
\providecommand{\assumptionnumber}{}
\makeatletter
\newenvironment{assumption}[2]
 {%
  \renewcommand{\assumptionnumber}{Assumption #1}%
  \begin{assumption*}%
  \protected@edef\@currentlabel{#1}%
 }
 {%
  \end{assumption*}
 }
\makeatother

\newcommand{\E}{\mathrm{E}}

\newcommand{\possessivecite}[1]{\citeauthor{#1}'s (\citeyear{#1})} 

\begin{document}
 \begin{singlespace}
\maketitle  
\thispagestyle{empty}
 \end{singlespace}

\begin{abstract}  
\noindent This paper investigates the impact of Reconstruction-era amnesty policy on the officeholding and wealth of elites in the postbellum South. Amnesty policy restricted the political and economic rights of Southern elites for nearly three years during Reconstruction. I estimate the effect of being excluded from amnesty on elites' future wealth and political power using a regression discontinuity design that compares individuals just above and below a wealth threshold that determined exclusion from amnesty. Results on a sample of Reconstruction convention delegates show that exclusion from amnesty significantly decreased the likelihood of ex-post officeholding. I find no evidence that exclusion impacted later census wealth for Reconstruction delegates or for a larger sample of known slaveholders who lived in the South in 1860. These findings are in line with previous studies evidencing both changes to the identity of the political elite, and the continuity of economic mobility among the planter elite across the Civil War and Reconstruction.
\end{abstract}	

\pagebreak
\setcounter{page}{1} 


\section{Introduction}

Did Reconstruction-era policies expand opportunities for political mobility for those outside of the Southern planter elite? \cite{woodward1981} and other revisionist historians argue that the planter class did not regain power following Reconstruction, but rather, political power gravitated toward urban professional upstarts and erstwhile Whigs. According to this view, the new Democratic elite rose to the top of Democratic parties and monopolized public offices in the South after 1876, and aligned itself with the economic interests of Northeast capital and against local agrarian interests. Economic institutions favoring elites persisted, despite policies allowing for more economic mobility among urban professionals.\footnote{\cite{woodward2003} argues that while some members of the planter class resumed farming operations despite land and crops at depressed values and lower labor productivity, more resourceful landholders ``transformed themselves into members of the new class that was creating a commercial revolution and fostering an industrial revolution.''}

There are few empirical studies that document changes to the identity of the Southern political elite in pre-and post-war periods. \possessivecite{cooper2005} study reveals that almost two-thirds of the South Carolina legislators in 1850 and 1860 were planters or farmers, while two-thirds of the sample were lawyers in the postbellum. This shift in the occupational backgrounds of Southern politicians is consistent with \possessivecite{woodward1981} assertion that railroad executives, corporation lawyers, and speculators comprised the political elite in the Democratic Party during the ``Redeemer'' period, from the 1870s to 1910 \citep{moore1978}.

Several studies examining historical census data or tax records show both continuity in elite persistence across the Civil War and greater opportunities for elite mobility. For example, \cite{wiener1976} finds ``virtually no evidence that the [Civil War] and Reconstruction led to a `revolution in land titles,' or to the `downfall of the old planter class'." The author estimates that 43\% of the wealthiest planter families living in five Alabama counties in 1860 remained in the elite a decade later --- a persistence rate not much different than the pre-war persistence rate of 47\%.\footnote{\cite{wiener1976} defines the planter elite as those having real estate holdings of at least \$10,000 in 1850, \$32,000 in 1860, and \$10,000 in 1870.}

Leveraging newly digitized full-count census data, several recent studies find substantially more turnover at the top of the wealth distribution in the South compared to the North between the 1860 and 1870 censuses. \cite{dupont2016}, for example, find that almost half of the top five percent of Southern property-owning heads of household in 1870 had been among the top property-holders in 1860; however, the rate of persistence in the South was considerably lower than it was in the North over the same period. The authors conclude that this pattern demonstrates both considerable persistence among wealthy Southerners and greater opportunities for economic mobility following the Civil War. \cite{NBERw25700} find that Southerners held at least 50 percent less wealth than Northerners in the same wealth percentile in 1870. The authors find that much of the loss in economic status incurred by wealthy slaveholders in the South following the Civil War was recovered by their families within a generation. 
 
Reconstruction-era presidential amnesty policy may have contributed to elite turnover. Amnesty policy granted near-universal amnesty to ex-Confederates in exchange for taking a loyalty oath. Fourteen classes of individuals, most notably those with taxable property exceeding \$20,000, were excluded from amnesty and instead had the option to apply for a presidential pardon. The \$20,000 threshold was chosen by President Andrew Johnson, with the intention that it would ``punish and impoverish'' the wealthiest slaveholders who he believed were most complicit in pushing for Southern succession \citep[Ch.~5]{foner2011}. Excluded individuals were denied property rights and could not vote or hold public office for nearly three years. Amnesty policy potentially contributed to the rise of urban professional Whigs in the Southern Democratic party because it denied voting rights for many Southern Democrats and former secessionists.\footnote{\cite{alexander1961} writes that when voting rights were restored, ``such a large proportion of the Southern Whigs was now within the Democratic and Conservative parties of the Southern states that the chief issue seems to have been division of the spoils between Whigs and Democrats, or even between factions of each party group."} Amnesty policy may have also had a deleterious effect on the wealth of the planter elite because excluded landholders could not reclaim property that was confiscated or abandoned during the Civil War.\footnote{However, all 430,140 acres of land seized under the Second Confiscation Act of 1862 were returned to its original owners by 1867. Source: House of Representatives 39th Congress 2d Session Executive Document Number 99.}

The \$20,000 exception to presidential amnesty motivates the use of a regression discontinuity (RD) design. Specifically, I compare individuals with taxable property wealth marginally below and above \$20,000 to estimate the effect of being excluded from amnesty on ex-post officeholding and future wealth. The key idea is that individuals with wealth just under \$20,000 are on average very similar to those with wealth just over \$20,000 in all aspects except for their treatment status. I estimate the effect of being excluded from general amnesty on two samples: known slaveholders who lived in the South in 1860 and delegates who served in Reconstruction conventions. I estimate about 13\% of slaveholders and a quarter of delegates were excluded from general amnesty in 1865 because they had prewar wealth over \$20,000. I show that delegates marginally above the \$20,000 threshold are much less likely to hold office after the conventions than delegates marginally below \$20,000. This finding is consistent with the hypothesis that presidential amnesty policy may have depressed elites' public service participation by causing resentment or cutting ties in political networks. I find no evidence of treatment effects on later census wealth for either sample, which is consistent with the idea of continuity of economic mobility for the Southern elite across the Civil War and Reconstruction.

The paper is closely related to the literature on the persistence economic institutions. \cite{acemoglu2006, acemoglu2008americas, acemoglu2008persistence}, for instance, theorize that elites offset the loss of monopoly \textit{de jure} political power due to political reforms by heavily investing in \textit{de facto} political power (in the form of capture of political parties, violence, or disenfranchisement) in order to preserve economic institutions that are favorable to the elite.  According to the ``iron law of oligarchy,'' the persistence economic institutions does not depend on the identity of the elites; when newcomers capture political parties dominated by the old elite, the new elite maintain similar economic institutions as the ousted elite. This theory explains how an economic system based on slavery and labor-intensive cotton production could be replaced by repressive tenant farming system even when there was substantial turnover in the dominant Democratic elite following the Civil War and Reconstruction.

This paper proceeds as follows: Section \ref{history} provides a historical overview of Reconstruction and presidential amnesty policy; Section \ref{research-strategy} describes the research strategy and procedures for estimating treatment effects; Section \ref{data} describes the data used in the study; Section \ref{results} provides empirical results on Reconstruction delegates and 1860 slaveholder samples; Section \ref{conclusion} concludes. 

\section{Historical background} \label{history}

Reconstruction refers to the period (1865--1877) during which state and federal laws were rewritten to guarantee basic rights for former slaves.\footnote{Figure \ref{chronology} provides a timeline of events during the Reconstruction era that are important to the present study.} It was also a time when biracial governments and constitutional conventions came to power in former Confederate states to codify Republican policies such as universal suffrage. 

An older historiography views Reconstruction as too ambitious and the exclusion of elites from politics as detrimental to the goals of Reconstruction, particularly because elites disproportionately faced the tax burden for Reconstruction government programs, such as free public education and railroad construction:
\begin{quote}
	\begin{singlespace}
	The result of all [these programs] was promptly seen in an expansion of state debts and an increase of taxation that to the property-owning class were appalling and ruinous \ldots [T]his class, which paid the taxes, was sharply divided politically from that which levied them, and was by the whole radical theory of the reconstruction to be indefinitely excluded from a determining voice in the government \citep[pp. 205-6]{dunning1907}.
\end{singlespace}
\end{quote}

A newer interpretation views Reconstruction as having not gone far enough. \cite{bensel1990}, for instance, argues that Reconstruction failed because Republicans chose neither massive redistribution of land from plantation owners to former slaves, or permanent military occupation of the South. \cite{foner2015} praises the Reconstruction Acts passed by Congress in 1867, which set up Reconstruction governments in the South, as a ``remarkable, unprecedented effort to build an interracial democracy on the ashes of slavery,'' and the ambitious public programs passed by these governments as evidence for the initial success of Reconstruction. Foner argues that Reconstruction did not last long enough, in part because Johnson too quickly abandoned the idea of depriving elites of their political and economic dominance.

\subsection{Presidential amnesty}\label{pres-amnesty}

The federal government passed several laws aimed at punishing and confiscating the property of Southerners who participated in the rebellion. Since it was seemingly unreasonable to apply laws of treason --- punishable by death --- to millions of Confederates, Congress responded to the rebellion by setting the terms of punishment for individuals who ``oppose[d] by force the authority of the government" to a fine of up to \$5,000 or imprisonment of up to six years.\footnote{U.S. Statutes at Large, 12: 284, (``An act to define and punish certain conspiracies"), approved July 31, 1861.} The First Confiscation Act permitted the confiscation of property, including slaves, used to support the rebellion. The Second Confiscation Act went further and allowed for the seizure of property from disloyal Southerners, and emancipated all slaves in territories under Union control.\footnote{U.S. Statutes at Large, 12: 319 (``An Act to confiscate Property used for Insurrectionary Purposes''), approved August 6, 1861. U.S. Statutes at Large, 12: 589-92 (An act to suppress insurrection, to punish treason and rebellion, to seize and confiscate the property of rebels, and for other purposes), approved July 17, 1862.} It also authorized the president ``to extend to persons who may have participated in the [the Confederacy] ... pardon and amnesty, with such exceptions and at such time and on such conditions as he may deem expedient for the public welfare.'' 

Acting upon Congressional authorization, President Abraham Lincoln issued two proclamations extending pardon to all individuals who took an oath to abide by the laws of the federal government --- with the exception of seven classes of individuals, including civil officials and military officers of the Confederacy --- allowing for members of the excluded classes to apply for a special pardon.\footnote{Proclamation 108, ``Amnesty and Reconstruction'', December 8, 1863; and Proclamation 111, ``Concerning Amnesty'', March 26, 1864.} The amnesty proclamation issued by Johnson in 1865 excluded fourteen classes of individuals, including Confederate state governors, and diplomatic or military officers of the Confederate government. The most expansive class of excluded individuals were those who had taxable property over \$20,000 (the 13$^{th}$ exception).\footnote{Proclamation 134, ``Granting Amnesty to Participants in the Rebellion, with Certain Exceptions,'' May 29, 1865.} According to \citep{fleming1905}, the assessment was made on the basis of 1861 taxable property:

\begin{quote}
\begin{singlespace}
	According to the proclamation the assessment was to be in 1865, but it was made on the basis of 1861 [taxable property], at which time slaves were included and a slaveholder of very moderate estate would be assessed at \$20,000. In 1865 there were very few people worth \$20,000.
\end{singlespace}
\end{quote}

Unlike Lincoln's proclamations, which held out no promise of amnesty to excluded classes, Johnson's proclamation declared that ``clemency will be liberally extended as may be consistent with the facts of the case.''\footnote{Johnson later remarked that he wanted excluded individuals to apply for a special pardon so they would ``realize the enormity of their crime" \citep[Chapter~6]{mckitrick1988}.} In most cases, excluded individuals had to apply first to their state's provisional governor, who then recommended applications for Johnson's administration to approve. Johnson approved virtually all applications received by his administration. Johnson granted 13,500 pardons by late 1867 \citep[Chapter~8]{dorris1953}, and in September of the same year, issued a second proclamation to reduce the number of excepted categories to three (the \$20,000 exception remained). Johnson's final amnesty proclamation issued on Christmas day in 1868 extended amnesty unconditionally to all those who participated in the rebellion.\footnote{Proclamation 167, ``Offering and Extending Full Pardon to All Persons Participating in the Late Rebellion,'' September 7, 1867; Proclamation 170, ``Granting Pardon to All Persons Participating in the Late Rebellion Except Those Under Indictment for Treason or Other Felony,'' July 4, 1868; Proclamation 179, ``Granting Full Pardon and Amnesty for the Offense of Treason Against the United States During the Late Civil War,'' December 25, 1868.} 

The policy goals of Johnson's amnesty policy is subject to historical debate. Foner speculates that the purpose of the policy was not to punish elites, but rather to force them to accept Reconstruction policies and to support the executive branch:
\begin{quote}
	\begin{singlespace}
		Johnson's policies had failed to create a new political leadership to replace the prewar ``slaveocracy''--- partly because the President himself had so quickly aligned with portions of the old elite
		\ldots 
		Why the President so quickly abandoned the idea of depriving the prewar elite of its political and economic hegemony has always been something of a mystery\ldots he came to view cooperation with the planters as indispensable to two interrelated goals --- white supremacy in the South and his own reelection as President \citep[pp. 433, 445]{foner2011}.
	\end{singlespace}
\end{quote}
Referencing Senator James Blaine's description of Secretary of State William Seward's views on Reconstruction, Foner further speculates Johnson sought to garner the support of elites who applied for a pardon. Blaine writes of the elites who were excluded from amnesty:
\begin{quote}
	\begin{singlespace}
		[I]t was [Seward]'s belief that they would more highly appreciate the benefit of amnesty by receiving it as an individual gift for which they were compelled to ask. Nor did [Seward] fail to see that the personal importance and prestige of the excluded classes were by the very fact of exclusion advanced in the South... By excluding these Southern leaders, their sense of self-importance was enhanced, their influence among their people was increased. Subsequently, by granting special pardon and amnesty to individuals of these excluded classes, as was intended from the first, [Seward] felt that he would be bringing each one to whom Executive clemency was extended under a sense of personal obligation to the President, and would thereby be increasing the influence of the Administration in directing the process and progress of reconstruction in the South \citep[p. 74]{blaine1886}. 
	\end{singlespace}
\end{quote}

\section{Research strategy} \label{research-strategy}

The basic idea for the RD design is to compare individuals with similar values of a continuous variable (the ``running variable) that determines treatment status in order to estimate treatment effects near a given threshold \citep{thistlethwaite1960}. The present study compares individuals marginally above and below a threshold of \$20,000 to estimate the effect of being excluded from amnesty. This strategy is similar to quasi-experimental studies in higher education that estimate the effects of financial aid receipt on student outcomes by exploiting an eligibility threshold in family income and assets \citep{kane2003,mealli2012}. It is also similar to evaluations of extended benefits on unemployment duration that use pre-unemployment income as the running variable \citep{lalive2008}.\footnote{These studies typically employ ``fuzzy'' designs, in which a value of the running variable falling above or below the threshold acts as an incentive to participate in treatment, but does not alone determine receipt of the treatment. For higher education evaluations, the assignment of financial aid may also be a function of student academic performance and application to receive aid. For unemployment evaluations, age start of unemployment and location may also receive receipt of extended benefits.} Most RD designs in political science use the margin of electoral victory as a running variable for identifying electoral effects under the assumption that close winners and losers are comparable \citep{caughey2011,eggers2015}. The proposed study may be the first to use individual wealth as a running variable.\footnote{\cite{eggers2009} and \cite{querubin2013} estimate the wealth returns to holding office using close-election RD designs. \citet{harvey2020applying} reviews opportunities for using RD designs to estimate policy effects relevant to American political development at the aggregate level (e.g., at the level of a municipality).}

Using \possessivecite{imbens2008} notation, which is situated within the potential outcomes framework \citep{neyman1923,rubin1990}, let $X_i$ be the running variable and $c$ be the cutoff value, which both determine treatment received:
\begin{align} 
W_i \in \{0,1\} =  \begin{cases}
1 & \mbox{if\,} X_i \geq c \nonumber \\
0 & \mbox{if\,} X_i < c. \nonumber
\end{cases} 
\end{align}  Let $Y_i(0)$ and $Y_i (1)$ be unit $i$'s potential outcomes in the absence and receipt of treatment, respectively. The observed outcome is thus $Y_i = (1-W_i) Y_i(0) + W_i Y_i(1)$. 

We cannot estimate the average treatment effect at $X_i=c$ because there are no control units at the cutoff, by definition. Instead, we observe units arbitrarily close to $c$, and assume that any association between the running variable and the potential outcomes is smooth:
\begin{assumption}{1}{}\label{a1}
Continuity: $\E [Y_i(0) | X_i = x]$ \text{ and } $\E [Y_i(1) | X_i = x]$ are continuous in $x$.
\end{assumption}
\noindent
Assumption \ref{a1} enables the interpretation of any discontinuity in the conditional expectation of the outcome given the running variable as evidence of a causal effect of the treatment. The Assumption is stronger than required because we're only interested in $x=c$, but as \citet{imbens2008} notes, it is reasonable to assume continuity for all values of the running variable. 

Given Assumption \ref{a1}, the RD estimand is the average treatment effect as $x$ converges to $c$:
\begin{align}
\tau_{ITT} &= \E [Y_i(1) - Y_i(0) | X_i = c] \nonumber \\
&= \lim_{x \downarrow c} \E [Y_i | X_i = x] - \lim_{x \uparrow c} \E [Y_i | X_i = x]. \label{srd}
\end{align}
The above estimand measures the average effect of intention-to-treat (ITT), or the effect on individuals assigned to treatment. However, we are really interested in estimating the effect of treatment-on-the-treated (TOT), or the effect on those actually treated. 

Let $\alpha$ denote the fraction of \textit{always--treats} in the study population --- those who accept treatment regardless of their assignment --- and $\gamma$ be the fraction of \textit{never--treats} who never accept treatment. Let $\beta$ be the fraction of compliers ---those who comply with their assignment --- and $\theta$ denote the fraction of \textit{defiers} --- those who behave contrary to their assignment. To estimate TOT effects, we assume that there are no always--treats and no defiers in the study population:
\begin{assumption}{2}{}\label{a2}
Single crossover: $\alpha = \theta = 0$ and $\beta > 0$.
\end{assumption} 
\noindent
There is no apparent threat to Assumption \ref{a2} because there are no \textit{always--treats} in the study population, by the definition of treatment as being excluded as a result of belonging to the \$20,000 class. \citet{freedman2006} shows that given Assumption \ref{a2}, the average effect of treatment on treated compliers is estimated by $\tau_{TOT} = \frac{\tau_{ITT}}{\beta}$. In the current application, compliers are excluded individuals who do not apply for a presidential pardon (virtually all special pardon requests were granted). \textit{Never-treats} are individuals would apply for a pardon if excluded for amnesty. Section \ref{data} describes the presidential pardon records, which I use to identify whether participants comply with treatment.  

I estimate the ITT effect using a local quadratic regression estimator proposed by \citet{calonico2014}. \cite{tukiainen2015} shows the this estimator recovers experimental results using random election outcomes, while the commonly-used local linear estimator \citep{imbens2011} does not recover the experimental benchmark.

\section{Data} \label{data}

In this section, I describe the Reconstruction delegates and 1860 slaveholders data samples, and presidential pardon records. 

\subsection{1860 slaveholders sample}\label{1860-slaveholders} 

I first draw a sample of over 1.25 million adult white Southerners from the full-count 1860 US Census\nocite{ipums, ancestry1860census}. The 1860 and 1870 censuses contain two measures of wealth: real estate and personal estate values.\footnote{Along with the 1850 Census, which asks for real estate but not personal estate wealth, the 1860 and 1870 Censuses are the only source for historical, nationally-representative data on individual wealth. Access to the full-count data is granted by agreement between UC Berkeley, and the Minnesota Population Center (IPUMS USA). The Minnesota Population Center has collected digitized census data for 1790-1930 microdata collection with contributions from Ancestry.com and FamilySearch.} The former captures the full value of real estate, even if the property was encumbered by debt. The latter captures all other forms of wealth, including cash, financial instruments, and the market value of slaves. I use the sum of real estate and personal estate values in each census year (``total census wealth'') as a proxy for locally-assessed taxable property. Census wealth is expected to be positively correlated with taxable property: \citet{steckel1994}, for example, matches households from the 1860 Census to state and local property records for samples in Massachusetts and Ohio and finds a strong linear relationship between total census wealth and taxable wealth. \citet{bleakley2016} find a strong linear relationship between 1870 wealth reported in county-level tax records and total census wealth in 1870.

The full-count census data does not have the wealth measures digitized, only information that is useful for genealogical inquiry, such as name, county and state of residence, age, gender, race, and birthplace. The full-count data also contains a URL link to the microfilm image for each record, so it is feasible to manually enter wealth data, at least for a subset of records. Since the majority of individuals in the \$20,000 class were slaveholders, I link the sample of adult white Southerners from the full-count 1860 Census to the 1-in-20 sample of the 1860 slave census \citep{slavepums}, which describes about 14,000 slaveholders following the procedure described in Section \ref{linkage}. I am able to successfully match 96\% of slaveholders in the 1-in-20 sample to the full-count census. I then link the matched slaveholders to a sample of adult white Southerners from the full-count 1870 Census in order to measure wealth in 1870. The match rate linking the matched sample to the 1870 census sample (18\%) is similar to the relevant match rate (20\%) reported in \citet{NBERw25700}. 

\begin{figure}[htbp]
	\begin{center}
		\includegraphics[width=\textwidth]{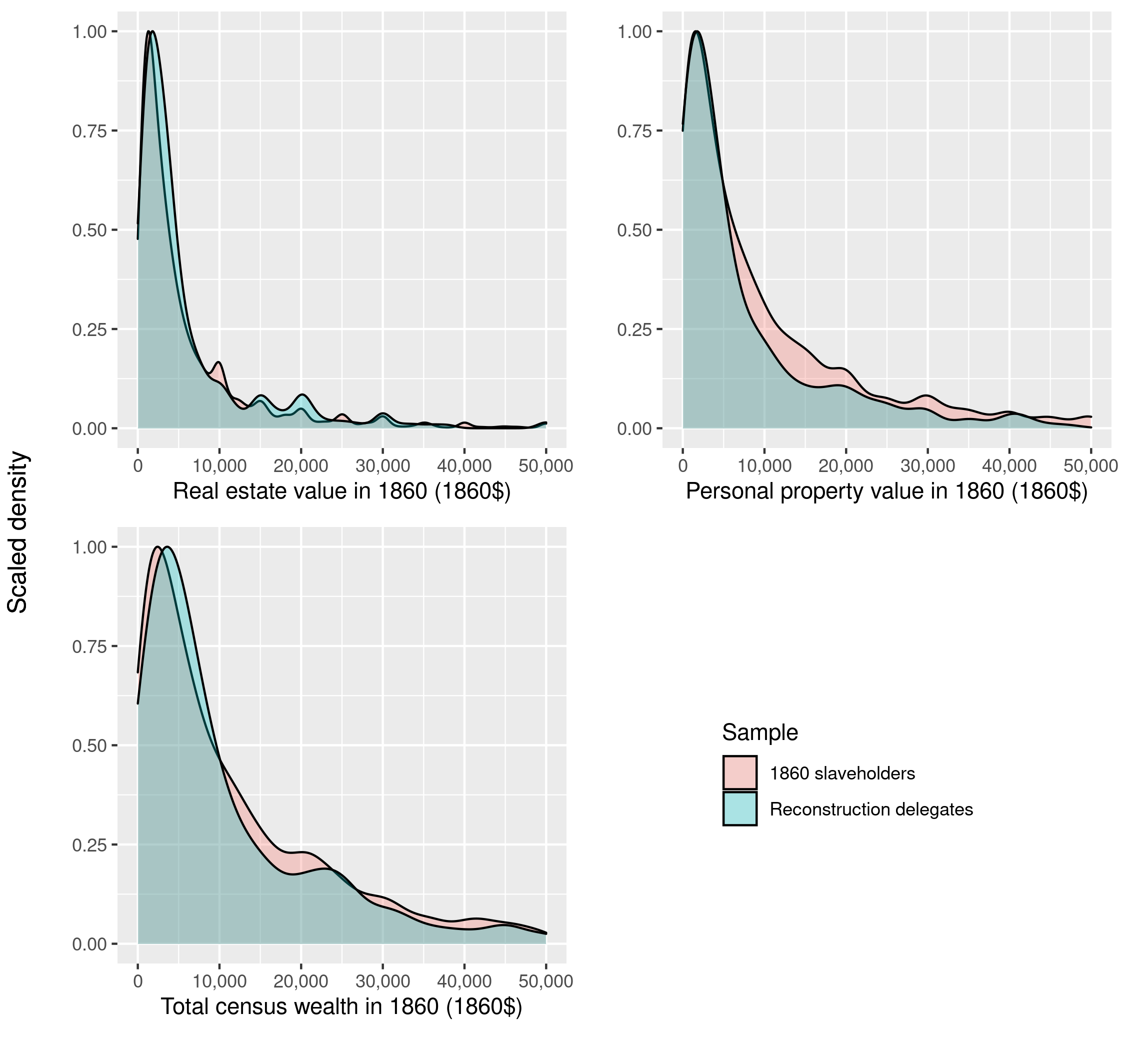} 
		\caption{1860 Census wealth densities for Reconstruction delegates and 1860 slaveholder samples.}
		\label{fig:wealth-plots}
	\end{center}
\end{figure}

I manually transcribe 1860 and 1870 census wealth information for over 5,000 slaveholders from the linked sample who lived in four Deep South states --- Alabama, Georgia, Mississippi, and South Carolina --- at the time of the 1860 Census. A potential issue with census wealth data is that missing values cannot be distinguished from zero wealth \citep{steckel1994}; I treat missing values as zero in order to preserve data. Figure \ref{fig:wealth-plots} shows that the distribution of 1860 Census wealth for this sample of slaveholders is comparable to the 1860 census wealth of a smaller sample of Reconstruction delegates described in Section \ref{delegates}.

Table \ref{data-sources} compares estimates on the number and percentage of individuals belonging to the \$20,000 exception. In the fully transcribed 1860 slaveholders sample, one quarter of those in the sample had total census wealth above \$20,000. It is not surprising that the estimated share of adult males belonging to the 13$^{th}$ exception is about twice the share in of a broader sub-population of adult males living in the South at the time of the 1860 Census: using the person-weighted 1\% sample of the 1860 Census, I estimate that 13\% of adult Southern males had over \$20,000 in total census wealth, which is not far off from estimates found in the historiography. \cite{dorris1953}, for example, estimates that up to 6\% of all Southerners belonged to the 13$^{th}$ exception, while \cite{avery1881} estimates that 10\% of adult males living in Georgia in 1860 were in the 13$^{th}$ exception. Both authors estimate up to 16\% were in all exceptions. 

\input{data-sources.tex}

Table \ref{slaveholders-sum} provides summary statistics on the transcribed 1860 slaveholders sample ($N=5,194$). The median total census wealth of the sample is close to \$7,000, which is over twice the value of a comparable 1\% sample from the 1860 Census. Slaveholders in the sample lost an average of about \$2,400 in personal property value between 1860 and 1870, which reflects the loss of slave wealth between the two censuses. Nearly 70\% of slaveholders in the sample report farmer or planter as their occupation. 

\subsection{Reconstruction delegates sample} \label{delegates}

I draw a second sample of Southern white delegates who served in state Reconstruction conventions, collected by \citet{hume2008}. The purpose of the state conventions was to adopt a new state constitution under the rules of Reconstruction in order to be readmitted to the Union \citep[Chapter~3]{dunning1907}. Conventions were held in all former Confederate states (except Tennessee) from 1867 to 1869 and were attended by a total 1,018 delegates. The majority of delegates were Southern-born whites, except for conventions in Florida, Louisiana, and South Carolina, which were majority black. Elections were held before Johnson's granting full pardon and amnesty to all ex-Confederates (Proclamation 179), with the exception of Texas, which held a second session after the Proclamation was issued.\footnote{Texas held its first session prior to Proclamation 179 on June 1-August 31, 1868, and its second session on December 7, 1868-February 6, 1869.} Ideally, we would have a sample drawn from the population of potential political actors (lawyers, planters, merchants, etc.), rather than a sample of actual political actors elected to the Reconstruction conventions. The advantage of the latter, however, is that ex-post officeholding is a likely outcome rather than a rare event. 

There are four types of delegates in the sample: (1) delegates who received amnesty; (2) delegates who were excluded from general amnesty, but received a pardon; (3) excluded delegates whose pardon status is not known; and (4) excluded delegates who did not receive a pardon. While only individuals who had received amnesty were allowed to participate as electors or delegates to the conventions, not all elected delegates had actually received amnesty or had been pardoned. \citet{dorris1953}, for example, claims that Johnson promised to provisional North Carolina Governor William Holden that elected delegates in North Carolina would receive a pardon, regardless of their prior status.\footnote{Holden appeared to favor prewar secessionists and his political supporters when recommending individuals for a presidential pardon.} I can account for each class of delegate in the sample, apart from the excluded delegates with unknown pardon status. As explained in Section \ref{pardon-data}, the pardon records data used in this study covers up to May of 1866, more than a year before the date of the first Reconstruction convention. 

I use delegate-level census and biographical data to create response variables on ex-post officeholding, future wealth, and ideology. First, I form a binary response variable that assumes unity if a delegate holds office after the convention. Second, I use census wealth in 1870 to assess the impact of not receiving amnesty on delegates' 1870 census wealth. Third, I create two measures of delegate ideology from the \citet{hume2008} data. I form a binary response variable that takes the value of one if the delegate formally protested the adoption of the constitution. I also use the authors' Republican support score that captures the share of votes each delegate cast with Republicans on five separate issue areas: economic issues (e.g., public spending on railroads); structure of the state government; racial issues (e.g., integration of public schools); suffrage issues; and state-specific issues (e.g., whether to partition the state of Texas). The Republican support score is calculated by taking the mean of pro-Republican votes within each issue area and then summing the score across issue areas for delegates with recorded scores in at least three issue areas. The score ranges from 0 (complete opposition of Republicanism) to 1 (complete support) for delegates voting in at least half of the roll calls in an issue area.

Table \ref{delegates-sum} provides summary statistics on the sample of Reconstruction delegates ($N = 574$). The median total census wealth of the sample is \$6,000, which is over twice the value of a comparable sample from the 1\% 1860 Census. About 23\% of the sample had total census wealth above \$20,000, compared to 13\% in the 1\% sample (Table \ref{data-sources}). On average, delegates lost over \$3,000 in personal property value between 1860 and 1870, which reflects the loss of slave wealth between the two censuses. About half of the sample were former slaveholders and the sample average for number of slaves held is eight (not shown in table). About a quarter of the sample previously held office, 12\% were noted Unionists, and 10\% were either a former Whig/Democrat or former Confederate. Support for Republican policies is evenly split and almost a third of delegates protested the adoption of the constitution.

The literature has not resolved whether Unionists were predominantly slaveholding planters predisposed to Republicanism or poor whites disadvantaged by the ruling planter class \citep[e.g.,][]{donald1944,trelease1963,ellem1972}. Figure \ref{fig:rss-wealth} shows that for the sample of Reconstruction delegates, support for Republicanism is decreasing in prewar total census wealth. For the median of each category, Figure \ref{fig:bin-wealth} shows that Southern unionist delegates tend be poorer than their ex-Confederate or Democratic counterparts. 

\subsection{Pardon applications}\label{pardon-data}

I identify individuals who applied for a special pardon in order to correct for imperfect compliance in the RD design. I rely on \possessivecite{douthat1999} Congressional records of special pardons granted by Johnson as of May 4, 1866, which includes the applicants' name and the reason for exclusion.\footnote{These records are found in U.S. House of Representatives, 39th Congress 2d Session Executive Document Number 99. More extensive records can be found in \cite{rowe1996}, which include both individual pardon applications to Johnson (1865 to 1867) and individual applications to Congress (1868 to 1898).} I link the 1860 slaveholders and Reconstruction delegates samples to pardon records following the procedure described in Section \ref{linkage}. I estimate that about 11\% of 1860 slaveholders in the \$20,000 class and 7\% of Reconstruction delegates in the \$20,000 class received a presidential pardon. The share of 1860 slaveholders in the \$20,000 class is likely understated, because the presidential pardon records used in this study not capture pardons issued between May of 1866 and the granting of general amnesty in 1868. Similarly, more Reconstruction delegates could have been granted a pardon before elections to the conventions, which were held in the months after Congress passed the Reconstruction acts in March of 1867. 

\section{Empirical results} \label{results} 

\begin{figure}[htb] 
	\begin{center}
		\includegraphics[width=0.49\textwidth]{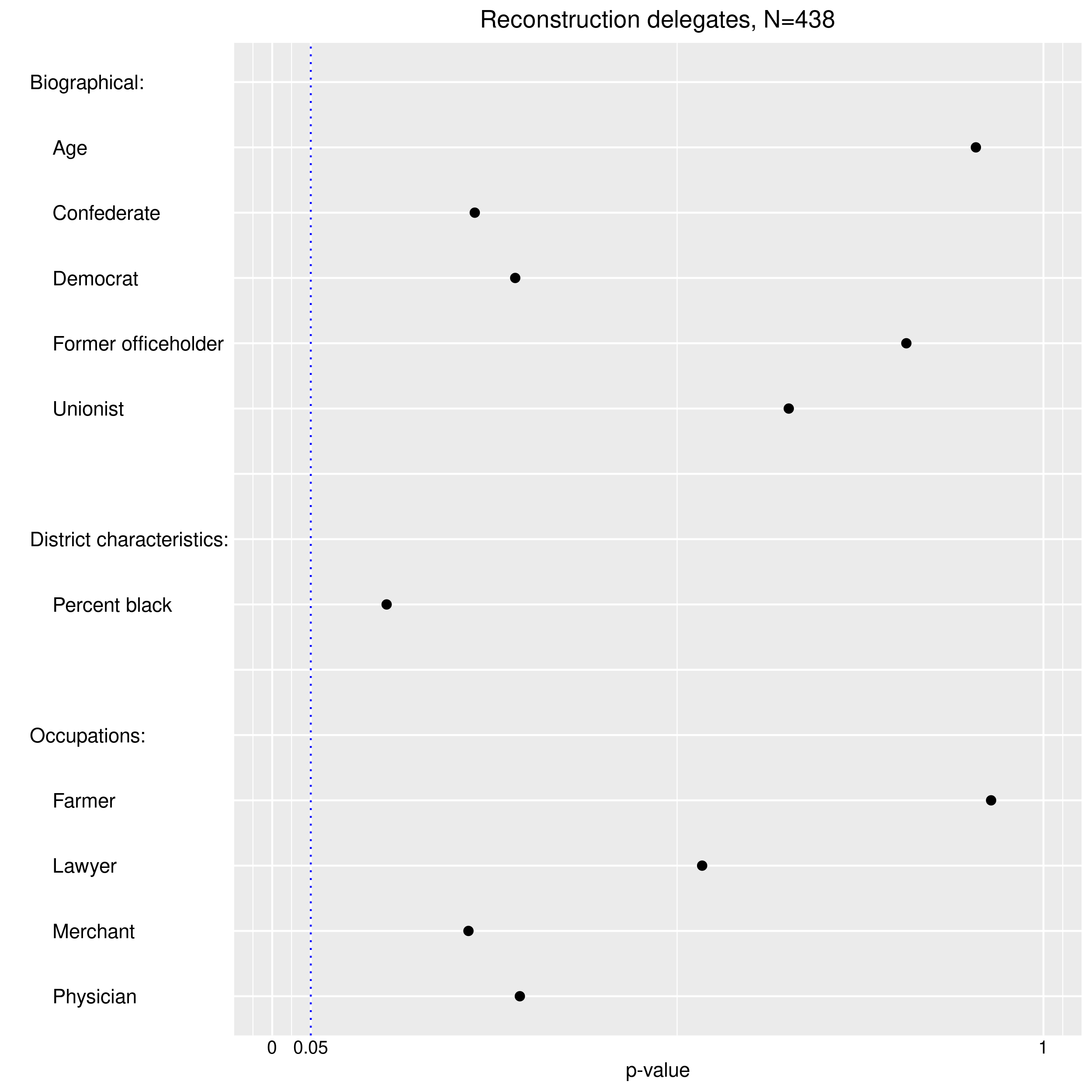} 
		\includegraphics[width=0.49\textwidth]{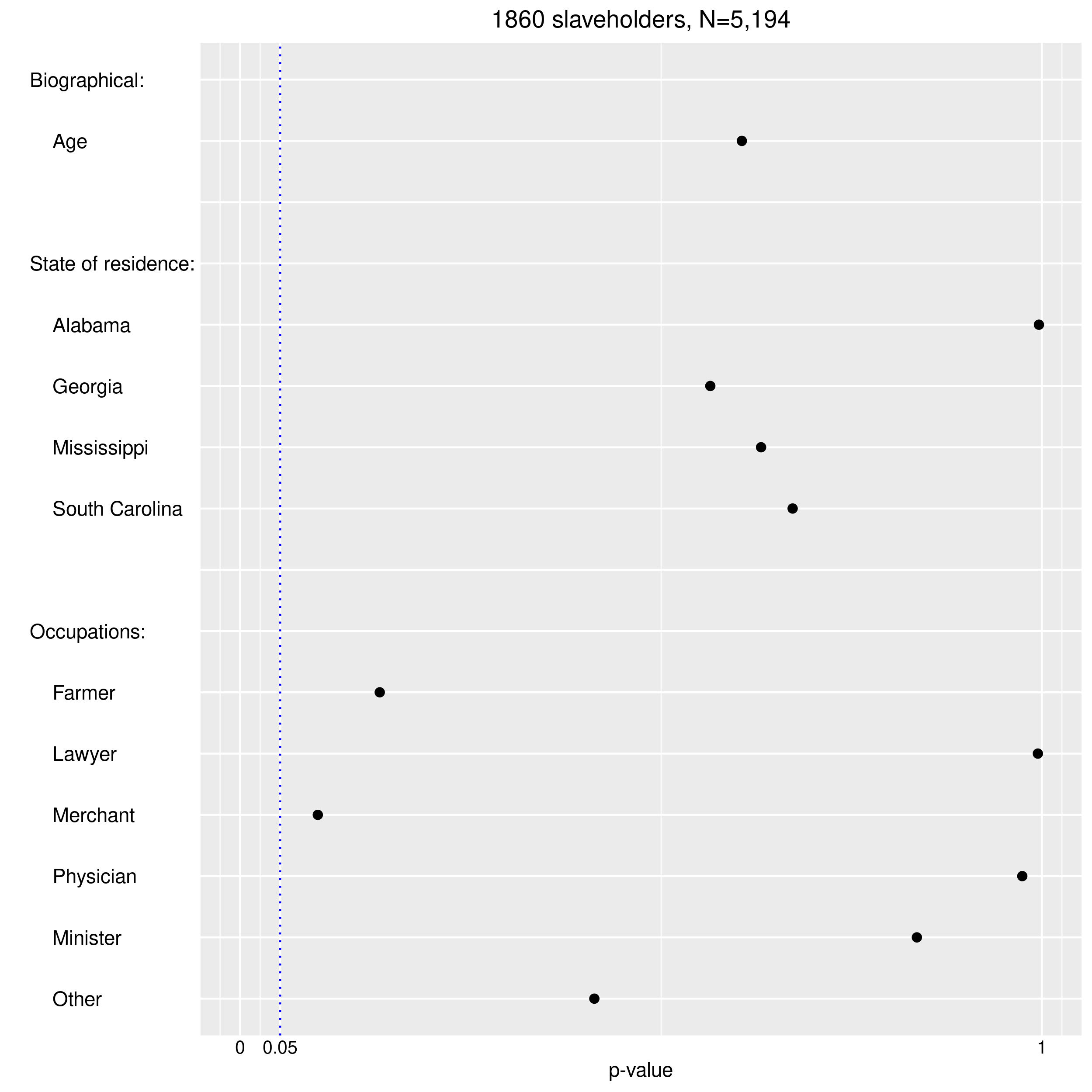} 
		\caption{Balance in treatment assignment for Reconstruction delegates and 1860 slaveholders samples. The dotted vertical line indicates the standard significance level.}
		\label{fig:rd-balance}
	\end{center}
\end{figure}

Prior to providing estimates of the effect of belonging to the \$20,000 class on the Reconstruction delegates and 1860 slaveholder samples, I first empirically test the key identifying assumption of continuity around the \$20,000 cutoff (Assumption \ref{a1}). \citet{caughey2011} recommend two tests of Assumption \ref{a1} that involve measuring balance in treatment assignment in terms of pretreatment covariates, or the density of control and treated units at various cutoff points. While Assumption \ref{a1} cannot be directly tested, if treatment assignment is random, we expect the pretreatment characteristics not to differ systematically based on treatment assignment. To test balance in treatment assignment, I consider each pretreatment covariate as an outcome and use the same estimation procedure for estimating ITT effects, described in Section \ref{research-strategy}. Figure \ref{fig:rd-balance} shows no significant difference between treatment and control groups in terms of pretreatment covariates such as age and occupation in either sample.

\begin{figure}[htb] 
	\begin{center}
		\includegraphics[width=0.49\textwidth]{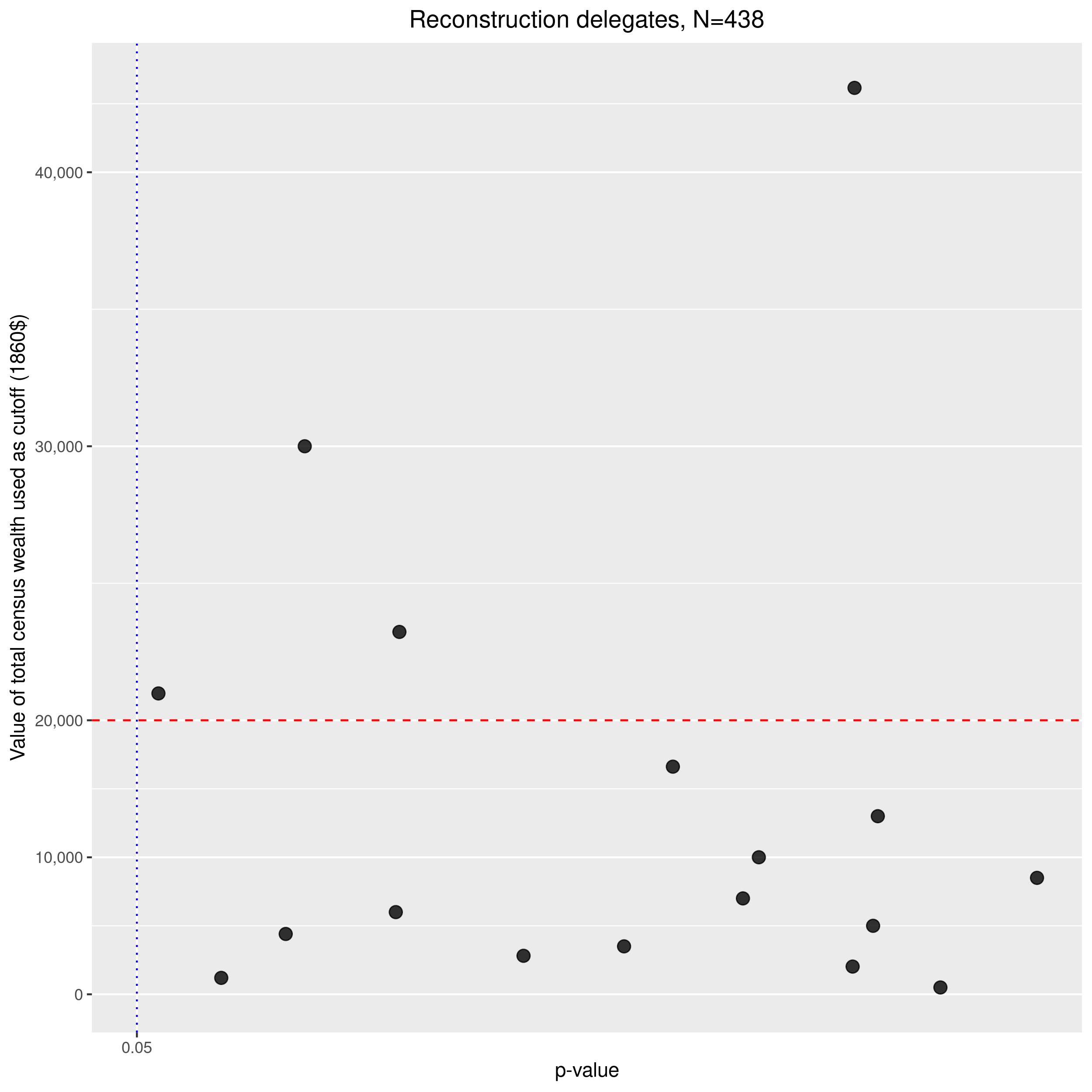}
		\includegraphics[width=0.49\textwidth]{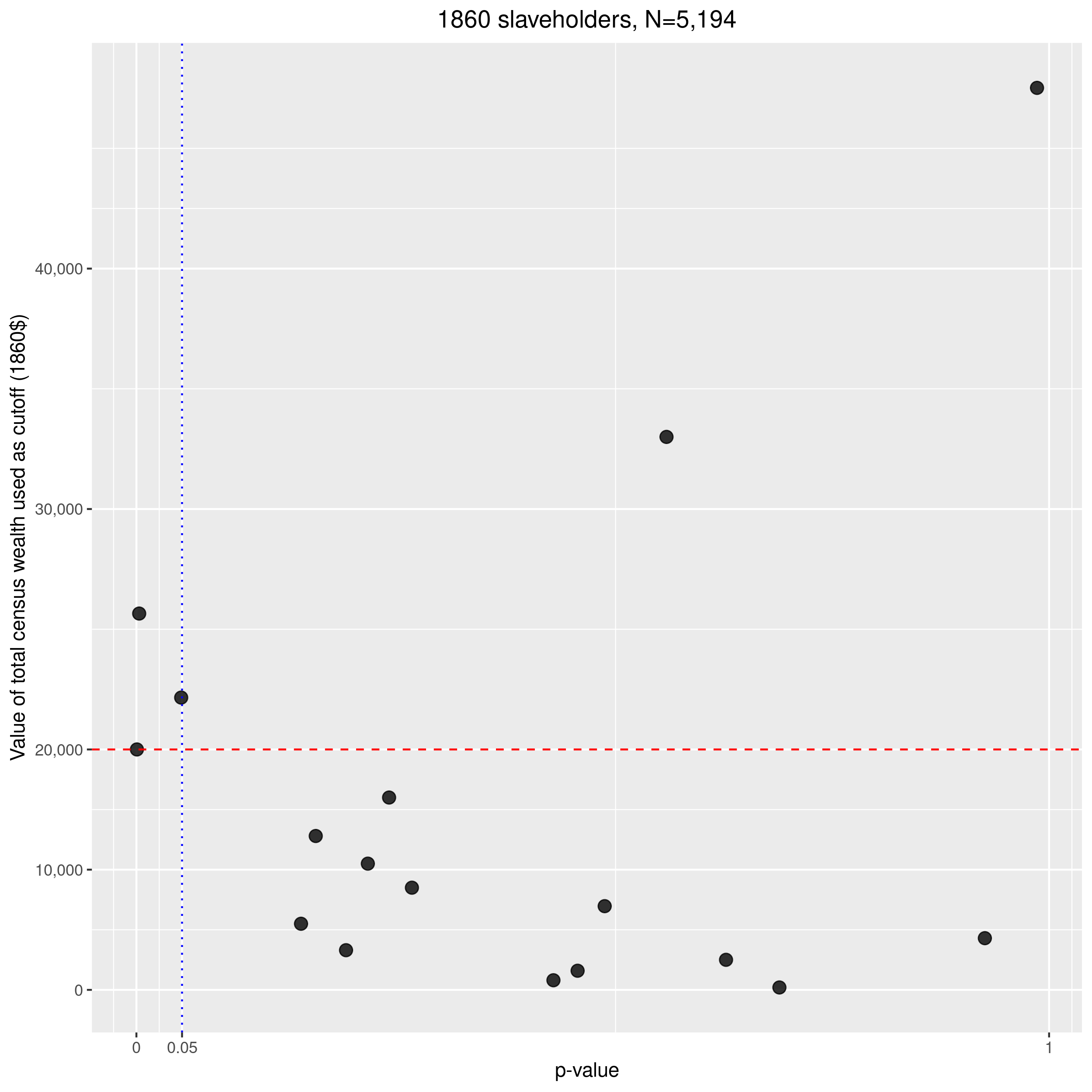} 
		\caption{Manipulation test $p$ values across possible cutoff values on Reconstruction delegates and 1860 slaveholders samples. The dashed horizontal line indicates the actual value used for the cutoff. The dotted vertical line indicates the standard significance level.}
		\label{fig:density}
	\end{center}
\end{figure}

The continuity assumption requires that units do not strategically select their treatment assignment. I test for a discontinuity in the density of control and treated units across a range of cutoff points using local polynomial density estimation \citep{cattaneo2015simple} and plot the results for each sample in Figure \ref{fig:density}. In the sample of Reconstruction delegates (left plot), I find no evidence of strategic manipulation for possible cutoff values ranging from \$200 to \$75,035. In the 1860 slaveholder sample (right plot), there is evidence of strategic manipulation at the actual \$20,000 cutoff ($p < 0.001$) and for the possible cutoff values \$24,314 ($p = 0.049$) and \$31,305 ($p = 0.003$). Since it is doubtful that slaveholders at the time of the 1860 Census could forecast the wealth cutoff for amnesty five years later, it is most likely that these significant manipulation results are detecting measurement error in the self-reported wealth. It should be underscored that self-reported census wealth data is generally considered in the historical political economy literature to be reliable.\footnote{See \citet{querubin2011}, \citet{bleakley2016}, and \cite{dupont2018economic} for a more comprehensive overview of the reliability of census wealth data. \cite{suryanarayan2021slavery} provide evidence of geographic variation in census reliability across Southern counties. Variation in census reliability would be problematic for the present study if it caused violations to the continuity assumption. While not directly testable, Fig. \ref{fig:rd-balance} demonstrates balance in treatment assignment in terms of state of residence for 1860 slaveholders and the percent of black residents in districts represented by Reconstruction delegates.}

\subsection{Main estimates}

Table \ref{rd-estimates-delegates} reports the causal estimates on binary and census wealth outcomes for the Reconstruction delegates sample. I find that delegates in the \$20,000 class are about 80\% less likely to hold office after the conventions, 95\% CI [-1.35, -0.30]. The treatment effect on treated compliers is even greater in magnitude, since it is obtained by scaling the ITT effect by the sample compliance rate, $\beta=0.93$. Controlling for noncompliance is important in this context because compliance status is not randomly assigned, and the delegates in the sample may behave differently than individuals in the \$20,000 class who complied with treatment and would have otherwise served in the convention.

I find no evidence of a treatment effect on the Republican support score, protesting the constitution, or 1870 census wealth.\footnote{Figures \ref{fig:pol-rd} and \ref{fig:wealth-rd} visually demonstrates a sizeable discontinuity at the cutoff point for the ex-post officeholder outcome, but not for the other response variables considered for the Reconstruction delegates sample.} The null effects may be explained by a lack of statistical power as a result of the small sample size of the Reconstruction delegates. I conduct \textit{a priori} power analyses by simulation following the procedures outlined in Section \ref{power} and present the results in Figure \ref{fig:power-indiv}. For census wealth outcomes, a treatment effect of well over $\$20,000$ is needed to achieve power over 80\% for a sample size of $N=400$, where 80\% is the power typically needed to justify a study. For binary outcomes, an (absolute) effect of over 70\% is needed for a sample size of $N=400$, which is in agreement with the statistically significant finding that treated delegates are about 80\% less likely to hold ex-post office.

\input{rd_estimates_delegates.tex}

Table \ref{rd-estimates-slaveholders} reports the causal estimates on census wealth outcomes for the 1860 slaveholders sample. I find no significant effect of being in the \$20,000 class on 1870 census wealth. The compliance-adjusted estimates are slightly larger, albeit nonsignificant, considering the sample compliance rate of $\beta=0.89$.\footnote{Relatedly, the regression discontinuity plots in Figure \ref{fig:wealth-rd} show no sizeable discontinuity at the cutoff point for the census wealth outcomes.} The null effects on the 1860 slaveholders same is unlikely to be explained by a lack of statistical power: the results of the power analysis (Figure \ref{fig:power-indiv}) shows that for census wealth outcomes, a treatment effect of $\$15,000$ is needed to achieve power over 80\% for a sample size of $N=5,000$. The TOT estimate on total census wealth for the sample of 1860 slaveholders carries a point estimate of more than $\$15,000$. 

\input{rd_estimates_slaveholders.tex}

\subsection{Subgroup analyses}\label{subgroup}

Two possible mechanisms underlying the estimated negative effect on ex-post officeholding for Reconstruction delegates is resentment due to being disenfranchised or losing connections in political networks as a result of being temporarily exiled. If the causal mechanism is losing connections in political network, I expect there to be larger effect sizes among previous officeholders because they are more embedded in political networks. I also expect amnesty policy to have a larger effect on Democrats and ex-Confederates than on Southern Unionists, since the latter were better positioned to reintegrate into post-Reconstruction politics. 

Table \ref{rd-estimates-delegates-subgroup} presents subgroup treatment effect estimates on ex-post officeholding among the Reconstruction delegates. In contrast to the hypothesis that presidential amnesty policy may have depressed elites’ public service participation by cutting ties in political networks, the estimated treatment effect among previous officeholders is smaller and non-significant. This finding implies that the significant treatment effect reported in Table \ref{rd-estimates-delegates} is driven primarily by Reconstruction delegates who did not previously hold office. Likewise, the estimated effect on Southern Unionists is smaller in magnitude and non-significant, which holds similar implications. However, the subgroup analyses are based on much smaller sample sizes and the estimates are consequently more difficult to interpret. 
 
\section{Conclusion} \label{conclusion}

The paper investigates the impact of Reconstruction-era amnesty policy on elite persistence. Individuals excluded from amnesty were disenfranchised and could not hold public office for nearly three years in which the policy was in effect. I use a natural experiment to investigate effect of being excluded from amnesty on the ex-post officeholding and future wealth of Southern elites.

Results on a sample of Southern white delegates to Reconstruction conventions show that being excluded from amnesty substantially decreased the likelihood of holding ex-post political office. Sub-group analyses imply this estimate is driven by Reconstruction delegates who did not previously hold office, and those who were not known to be ex-Confederates, Democrats, or Southern Unionists. I find no evidence that being excluded from amnesty diminished  delegates' support for Republicanism, adoption of the state constitution, or future wealth. Using a larger sample of known slaveholders in the 1860 Census, I find no evidence that being excluded from amnesty affected the future wealth of slaveholders. 

The results speak to the ``crucial question of continuity or discontinuity [of the Southern elite] across the Civil War" \citep{hackney1972} in the historical literature. The null findings on future census wealth for both samples are in agreement with previous studies evidencing the continuity of economic mobility for the Southern elite across the Civil War and Reconstruction. The finding that being excluded from general amnesty decreased ex-post officeholding for Reconstruction delegates is consistent with previous studies pointing to turnover in the political elite, as evidenced by the shift in occupational backgrounds of Southern politicians. Disenfranchisement might have facilitated the rise of a different class of political actors, thereby hastening the inevitable
change in the individual composition of the political elite. Together, the findings are in agreement with the theory that economic institutions can persist irrespective of the identity of the elites. Future research might explore how amnesty policy impacted elite persistence in other post-conflict democratization contexts; e.g., denazification in Germany and de-Ba'athification in Iraq. 

\newpage
\begin{singlespace}
\bibliographystyle{chicago}
\bibliography{references}
\end{singlespace}
	
\itemize
\end{document}


\begin{singlespacing}
\maketitle \thispagestyle{empty}
\tableofcontents \thispagestyle{empty}
\end{singlespacing}
	
\setcounter{figure}{0} \renewcommand{\thefigure}{SM-\arabic{figure}}
\setcounter{table}{0} \renewcommand{\thetable}{SM-\arabic{table}}
\setcounter{section}{0} \renewcommand{\thesection}{SM-\arabic{section}}
		
\pagenumbering{roman}
	
\pagebreak
\pagenumbering{arabic}

\section{Record linkage} \label{linkage}

I follow closely the procedure used by the Minnesota Population Center for linking microdata samples \citep{goeken2011,vick2011}. First, I preprocess the name fields by removing non-alphabetic characters and standardizing the most common first name strings. I then match records by the Soundex code of the surname, blocking on state. Second, I form a training set by manually labeling records as a correct or incorrect match in a subset (e.g., 10\%) of the matched records. Third, I fit an ensemble of algorithmic models on the training set, using Jaro-Winkler string distance in first and surnames and binary variables indicating exact first and surname matches as features of the model.\footnote{I evaluate the ensemble's performance estimating the 10--fold cross--validated risk on the training set.} Fourth, I use the ensemble fit on the unlabeled (test) data, using a prediction threshold to classify correct matches. Lastly, I exclude ambiguous multilinked records (e.g., multiple paired records for ``John Smith''). 

\section{Power analyses}\label{power}

The purpose of a power analysis simulation is to estimate $\mathrm{P}(\mathrm{Reject \, H_0} | \mathrm{H_0 \, is \, false})$ at a constant significance level of $\alpha =0.05$, for different treatment effects $\Delta_{1, \ldots, j}$ and sample sizes $N_{1, \ldots, k}$. I create a $j \times k$ sized grid that has each combination of $\Delta$ and $N$ as a cell in the grid. The simulation proceeds as follows:

\begin{enumerate}
\item Take a random sample of size $N_{jk}$ with replacement from a known distribution to create a vector of running variable values. I draw from the 1860 total census wealth values from the sample of known slaveholders in the 1860 Census described in Section \ref{1860-slaveholders}.
\item Simulate response values for treated and control units ---defined as units above or equal to and below, respectively, a threshold of \$20,000 in the sampled running variable--- with $\Delta_{jk}$ as the difference-in-means between treated and control units. I generate random values from the normal distribution for continuous responses ($\text{s.d.} = \$25,000$) and from the binomial distribution for binary responses ($\text{Prob.} = 0.1$).
\item Estimate Eq. \ref{srd} on the simulated data and extract the $p$ value.
\end{enumerate}
I repeat the simulation $\mathcal{L} =100$ times and calculate power of the test by dividing the count of the number of $p$ values that are less than $\alpha$ over $\mathcal{L}$. 

\clearpage
\section{Tables \& figures}

\begin{figure}[htbp] 
	\centering
	\input{chronology.tex}
	\caption{Chronology of events.}
	\label{chronology}
\end{figure}

\input{slaveholders-sum.tex}
\clearpage

\input{delegates-sum.tex}

\clearpage

\begin{figure}[htbp] 
	\begin{center}
		\includegraphics[width=\textwidth]{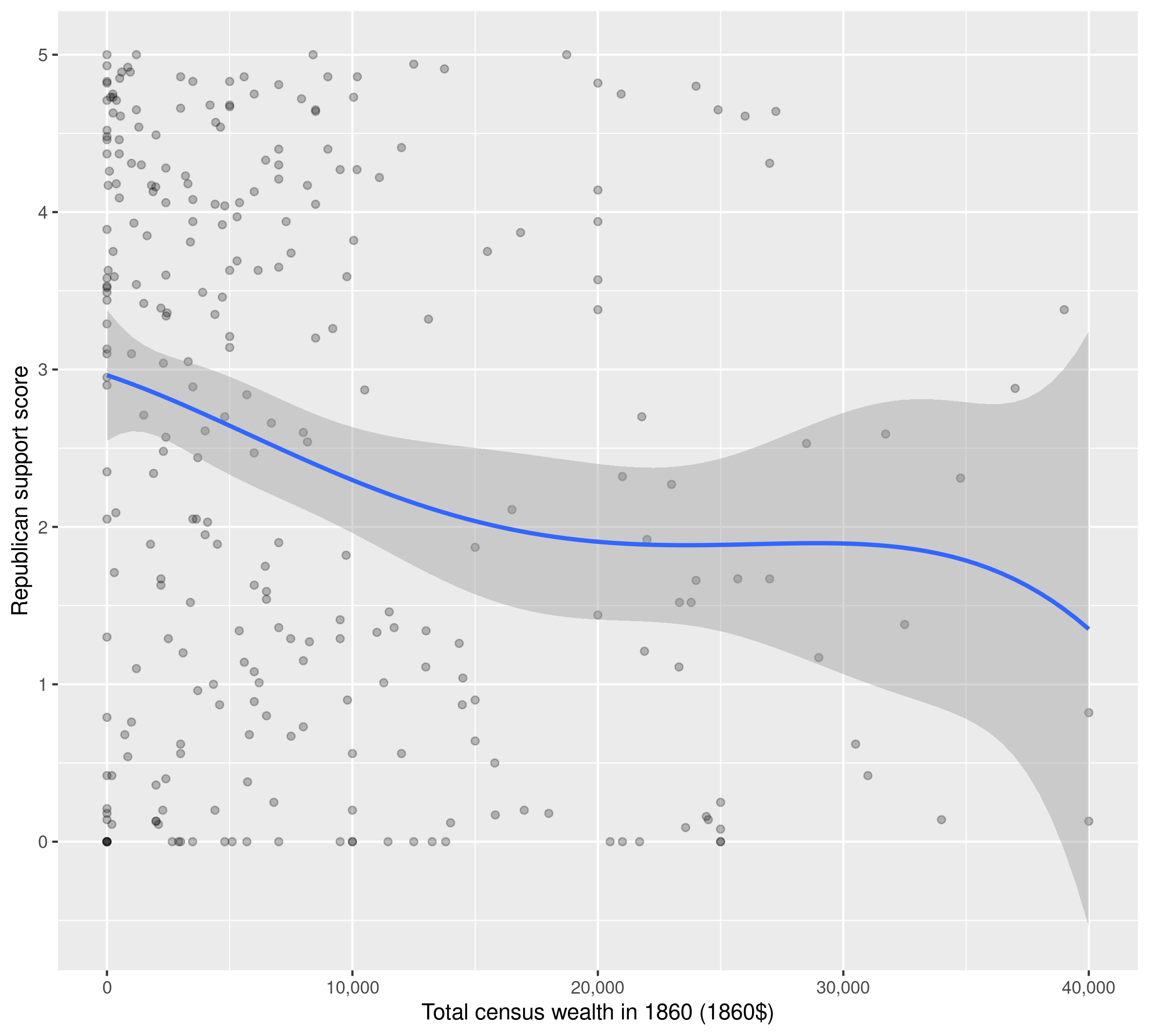} 
		\caption{Republican support score plotted against 1860 total census wealth (1860\$) for Reconstruction delegates with nonmissing census wealth ($N=438$). The solid line is the local fourth-order polynomial regression fit. The graph excludes wealth values greater than \$40,000.} 
		\label{fig:rss-wealth}
	\end{center}
\end{figure}

\begin{figure}[htbp] 
	\begin{center}
		\includegraphics[width=\textwidth]{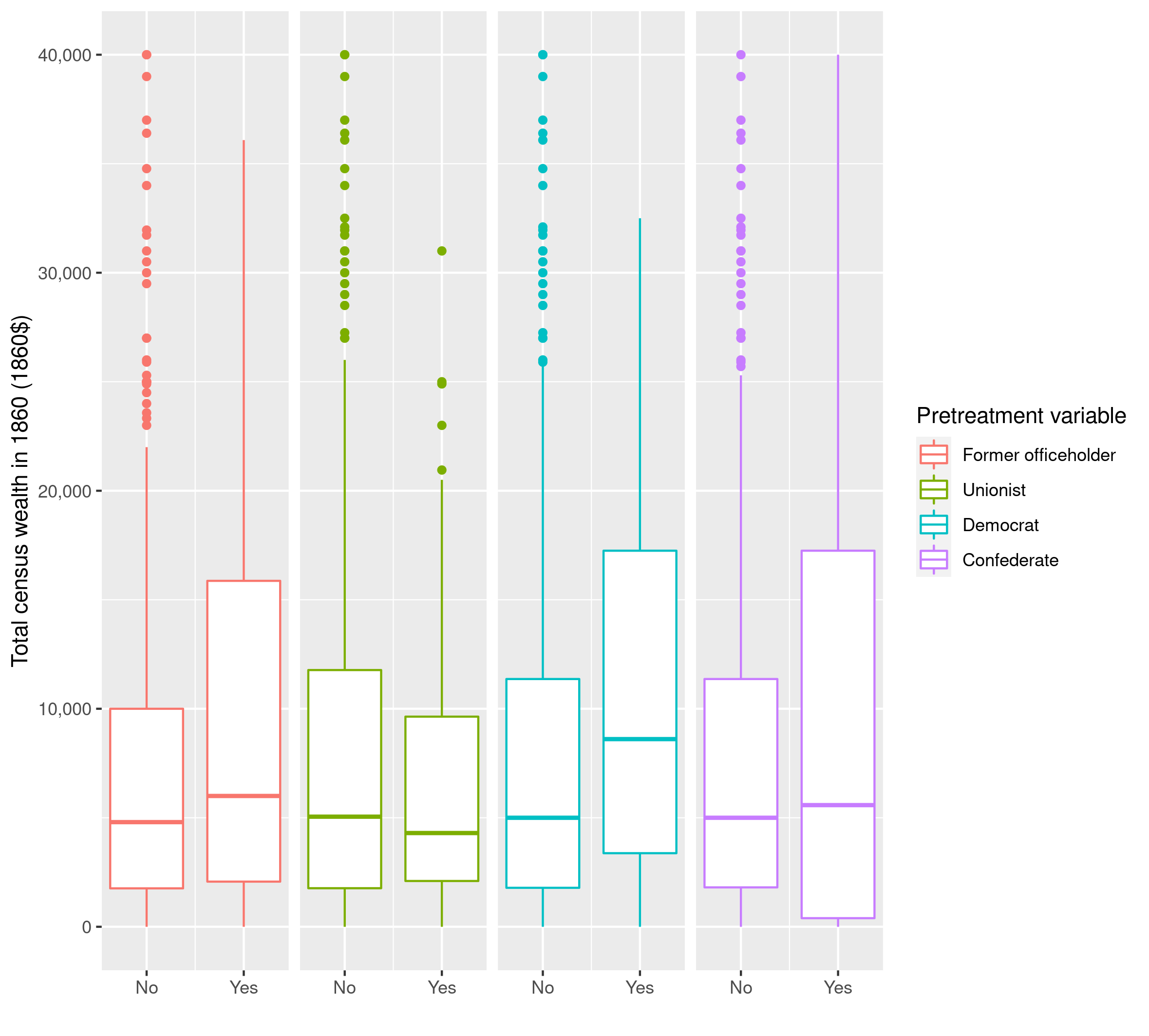} 
		\caption{Box plots for selected binary pretreatment variables, plotted according to 1860 total census wealth (1860\$), for Reconstruction delegates with nonmissing census wealth ($N=438$). The graph excludes wealth values greater than \$40,000.}
		\label{fig:bin-wealth}
	\end{center}
\end{figure}

\begin{figure}[htbp]
	\begin{center}
		\begin{tabular}{@{}cccc@{}}
			\includegraphics[width=.35\textwidth]{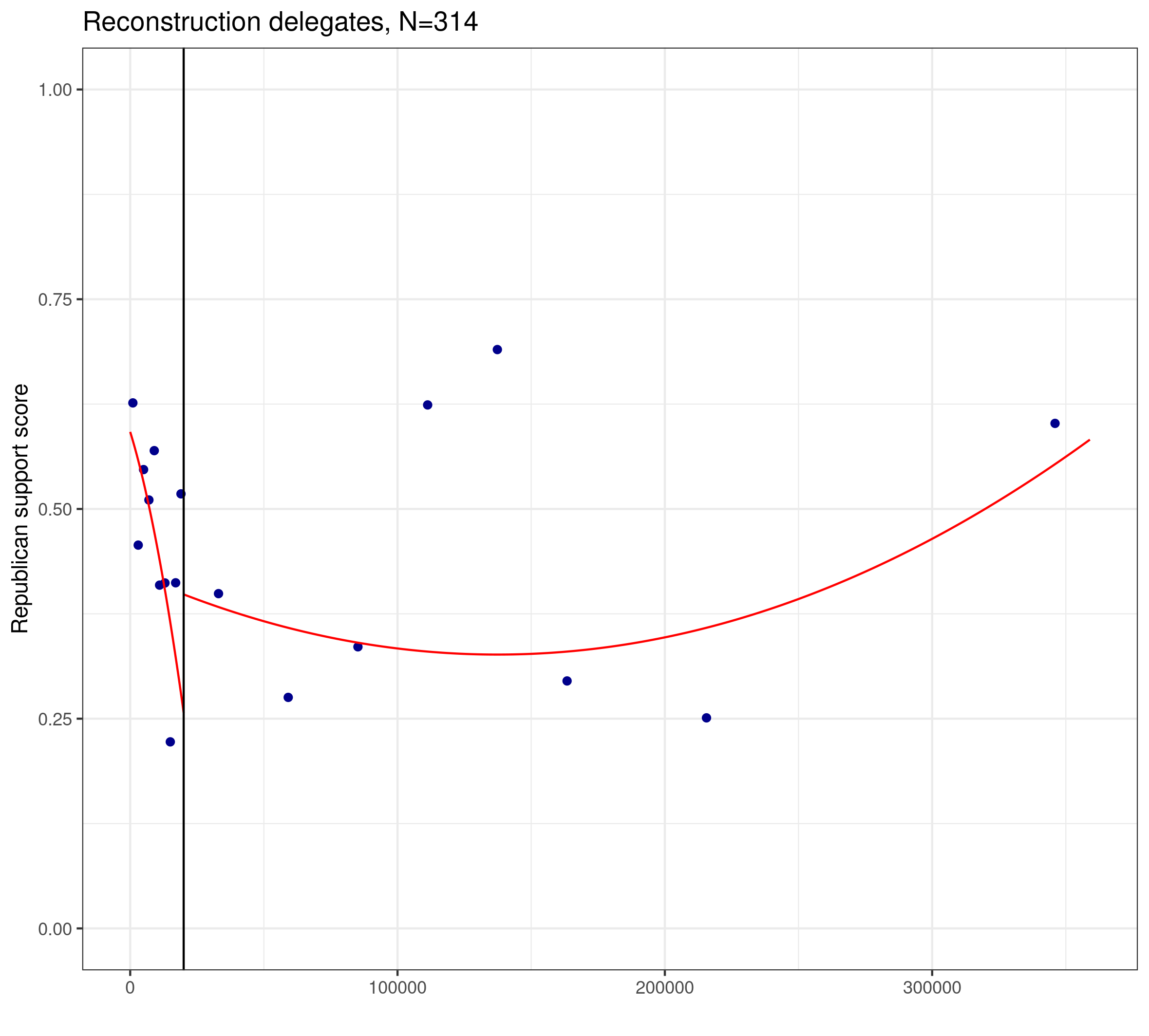}  &
			\includegraphics[width=.35\textwidth]{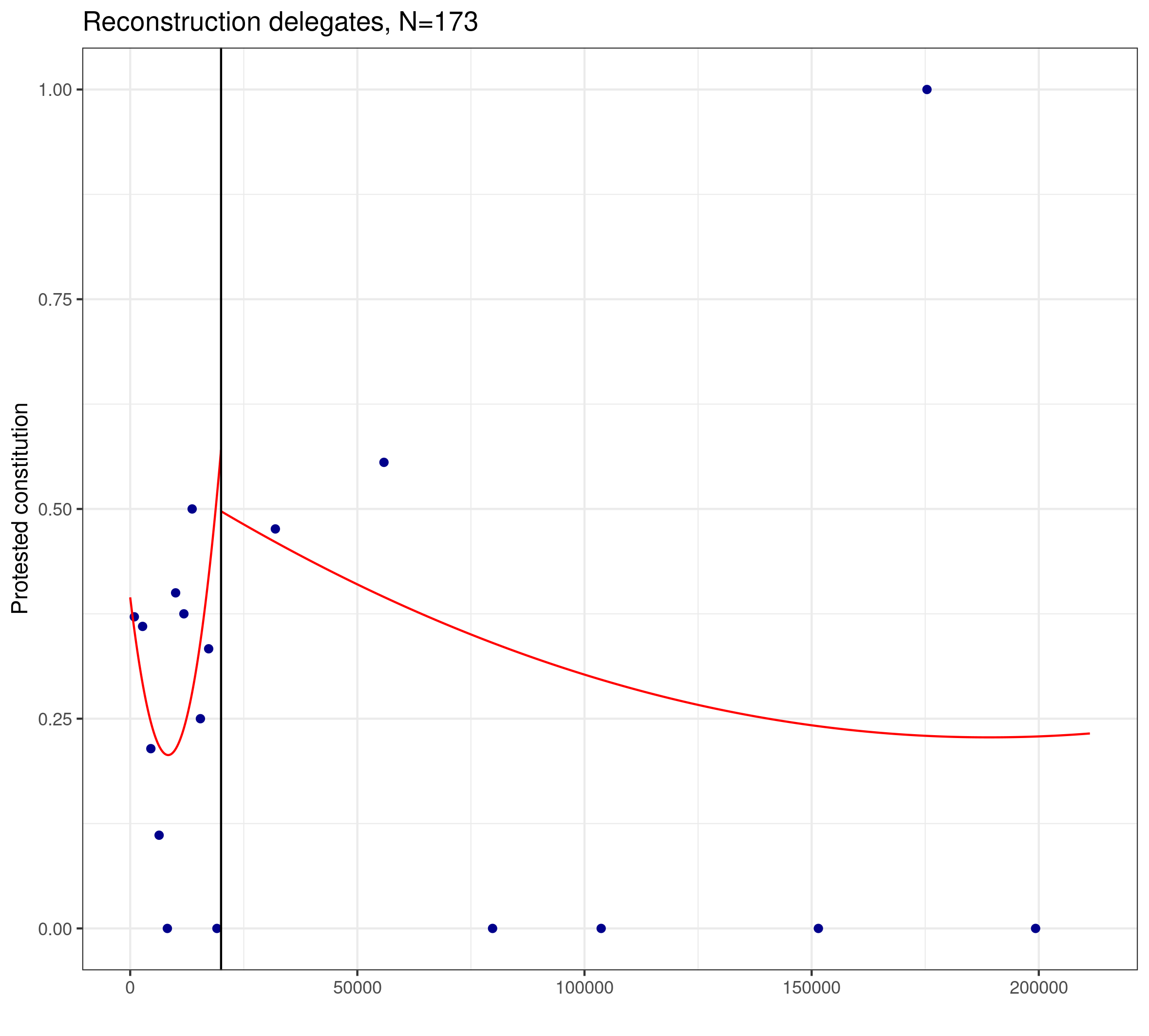} &
			\includegraphics[width=.35\textwidth]{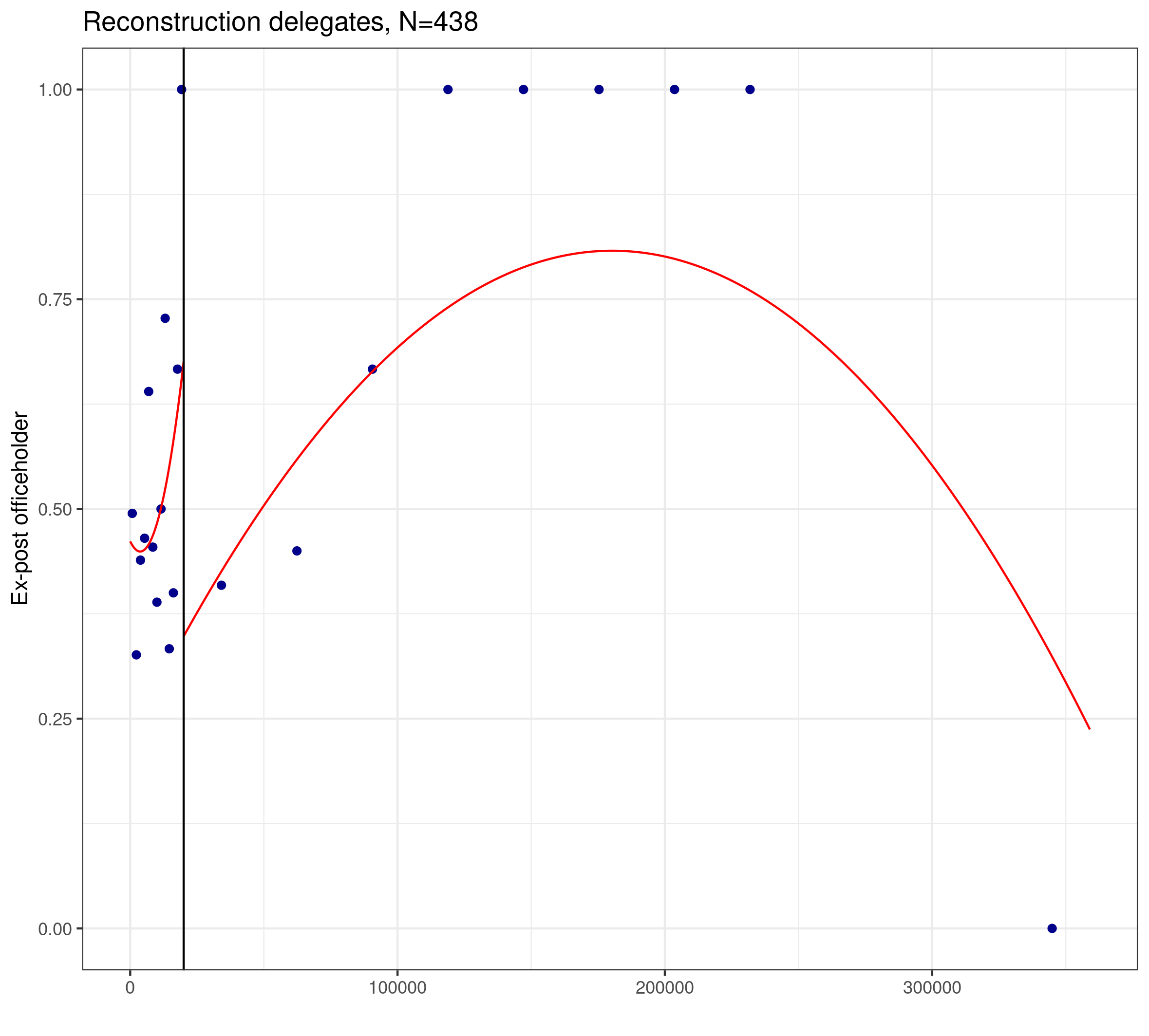} 
		\end{tabular}
		\caption{Delegates sample: regression discontinuity plots for binary responses, using binned sample means mimicking the underlying variability of the data. Curved lines depict second-order polynomial fits. The running variable (x-axis) is total census wealth in 1860 (1860\$) and the solid vertical line indicates the cutoff point of \$20,000.}
		\label{fig:pol-rd}
	\end{center}
\end{figure}

\begin{figure}[htbp] 
\begin{center}
  \begin{tabular}{@{}cccc@{}}
    \includegraphics[width=.35\textwidth]{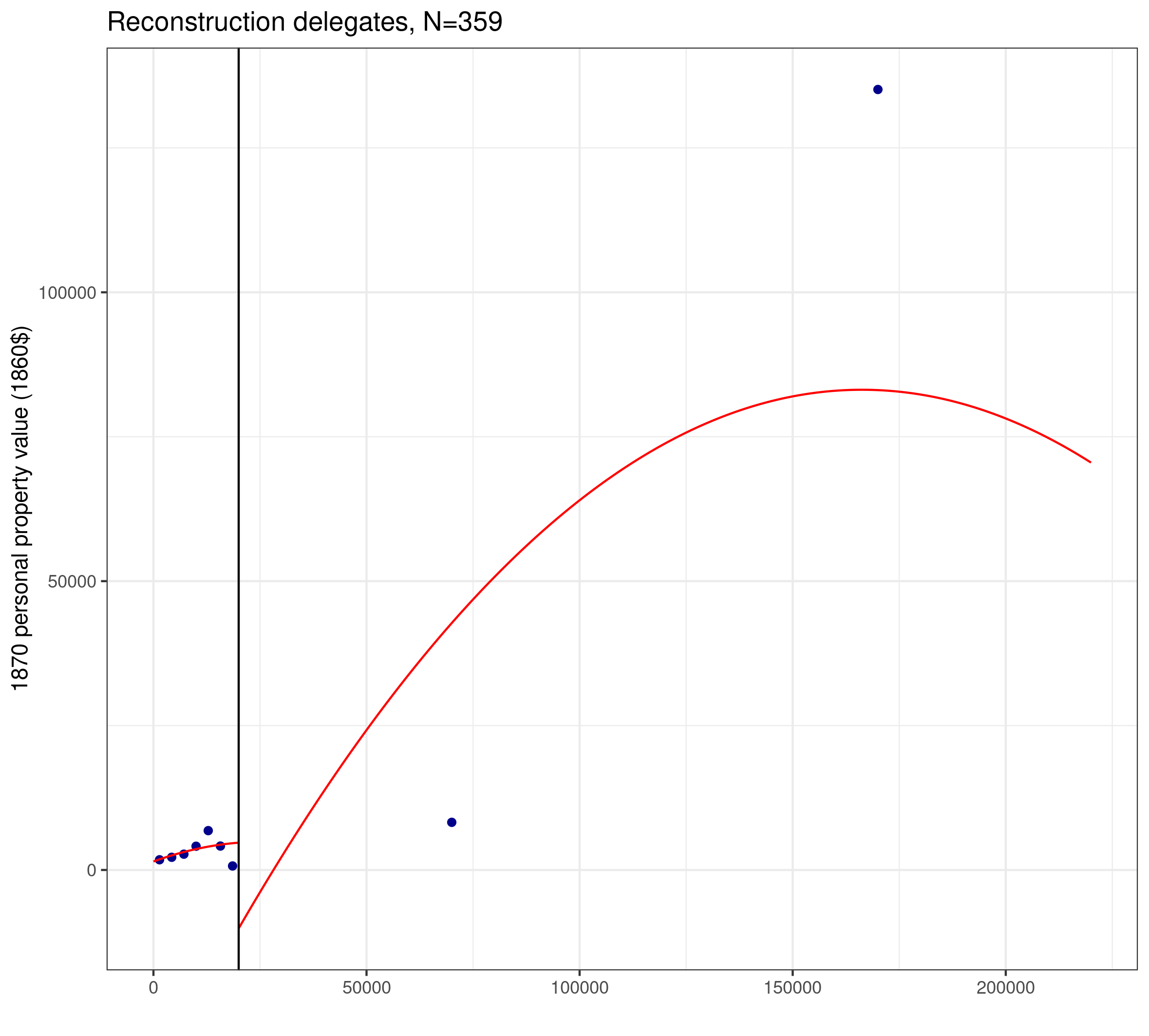} &
    \includegraphics[width=.35\textwidth]{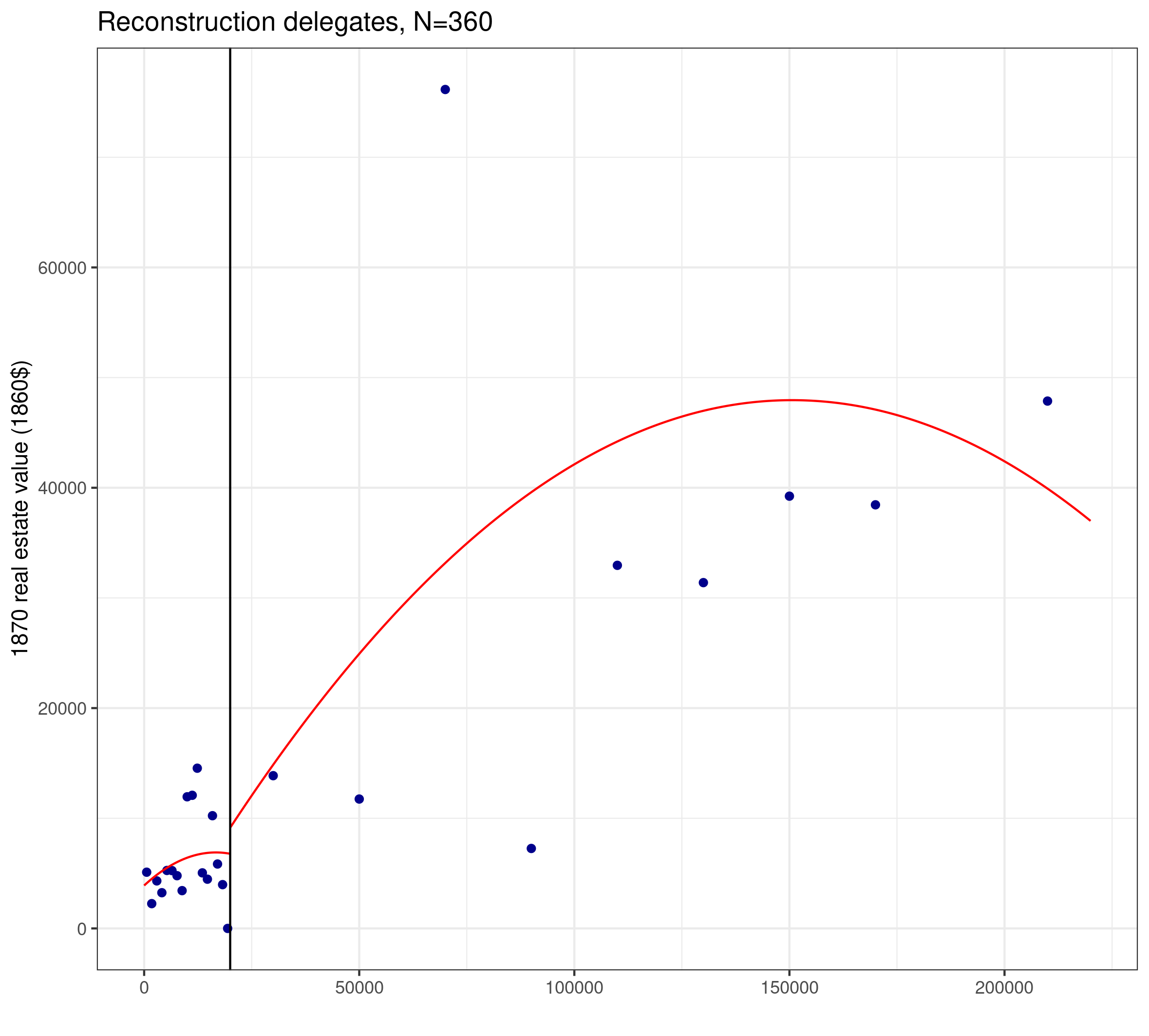} & 
   \includegraphics[width=.35\textwidth]{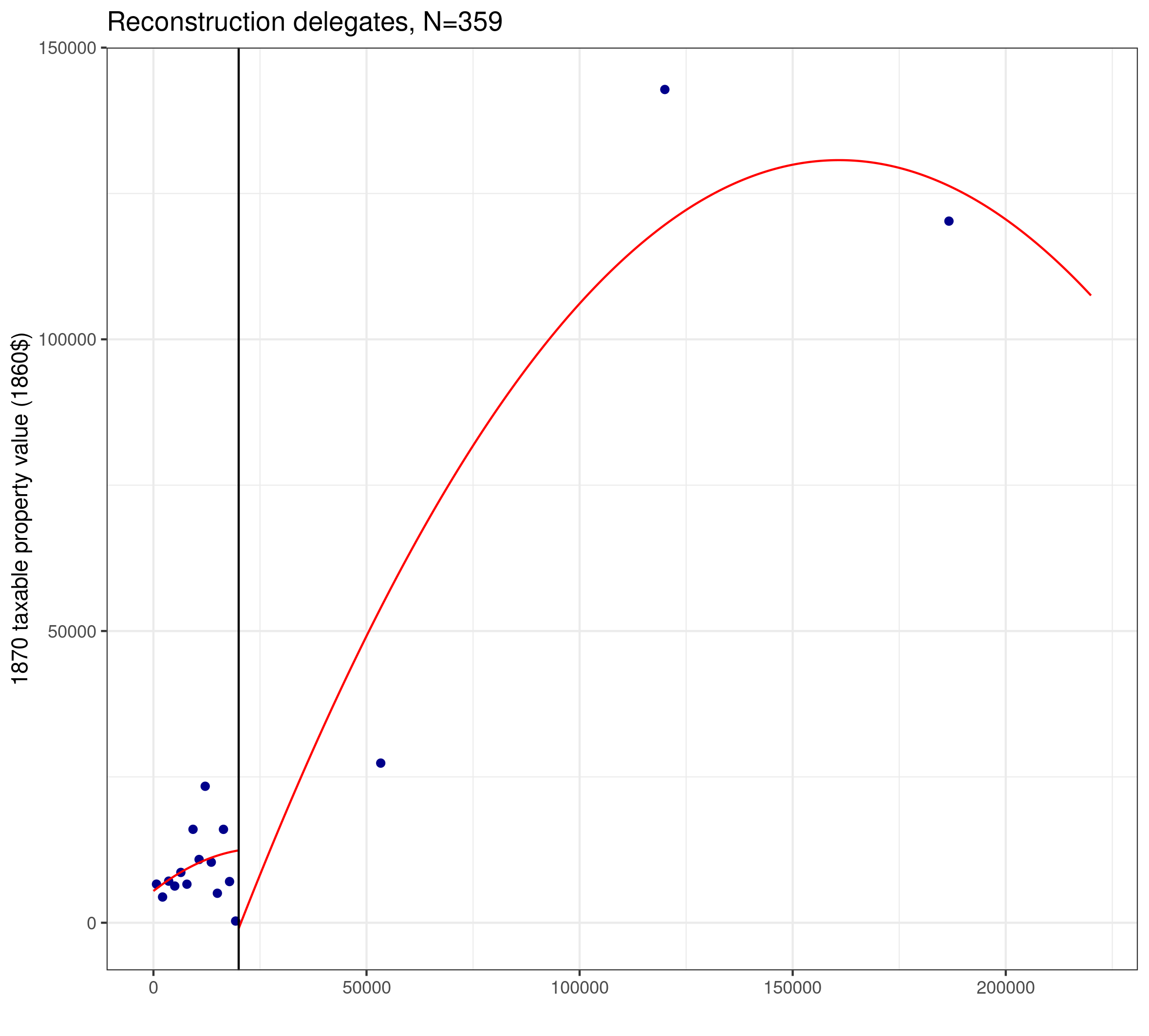} & \\
       \includegraphics[width=.35\textwidth]{plots/persprop_70.png} &
   \includegraphics[width=.35\textwidth]{plots/realprop_70.png} & 
   \includegraphics[width=.35\textwidth]{plots/taxprop_70.png} 
  \end{tabular}
  \caption{Reconstruction delegates sample: regression discontinuity plots for post-treatment wealth. See notes to Figure \ref{fig:pol-rd}.}
  \label{fig:wealth-rd}
  \end{center}
\end{figure}

\begin{figure}[htb] 
	\begin{center}
		\begin{tabular}{@{}cccc@{}}
			\includegraphics[width=.55\textwidth]{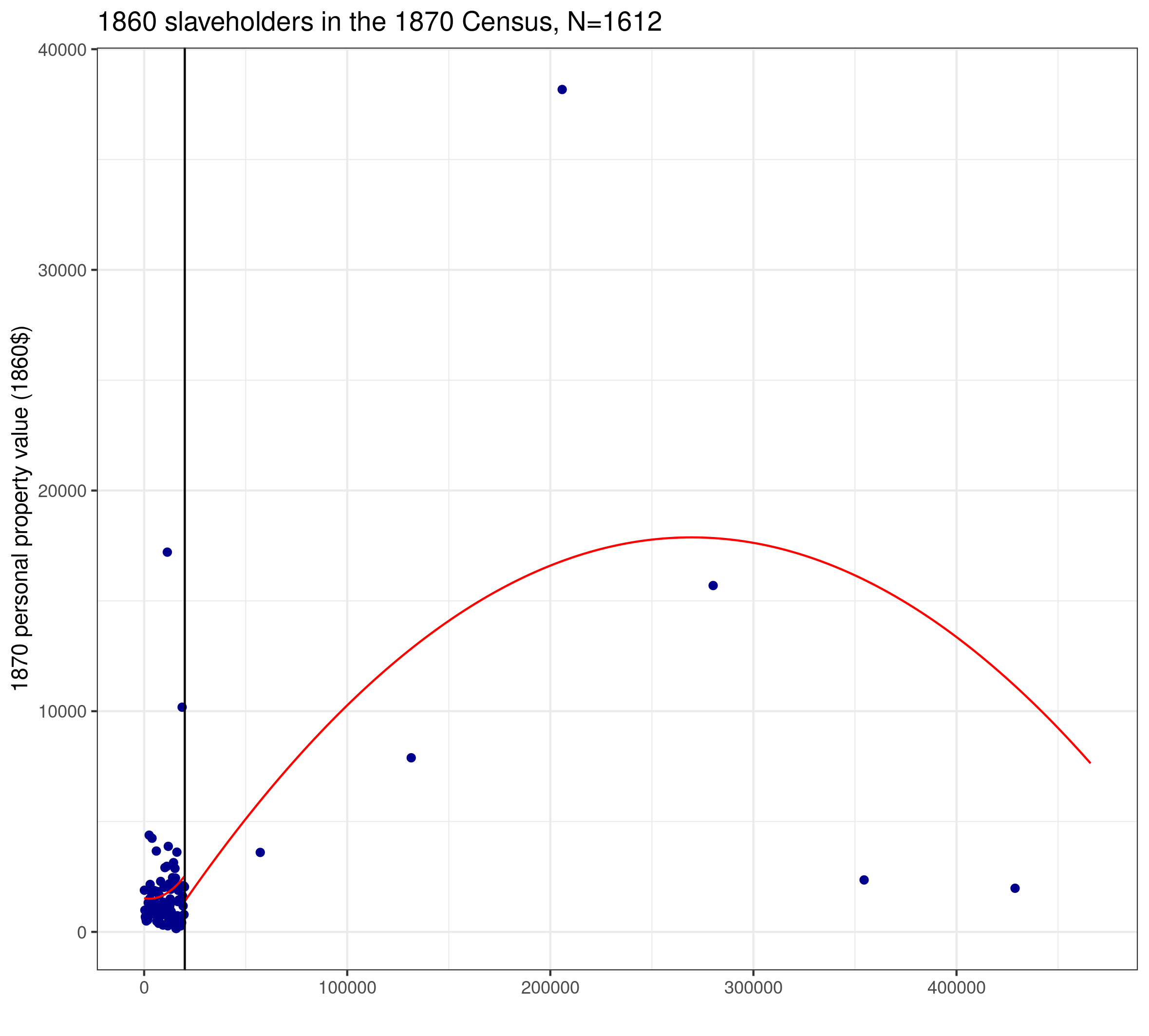} &
			\includegraphics[width=.55\textwidth]{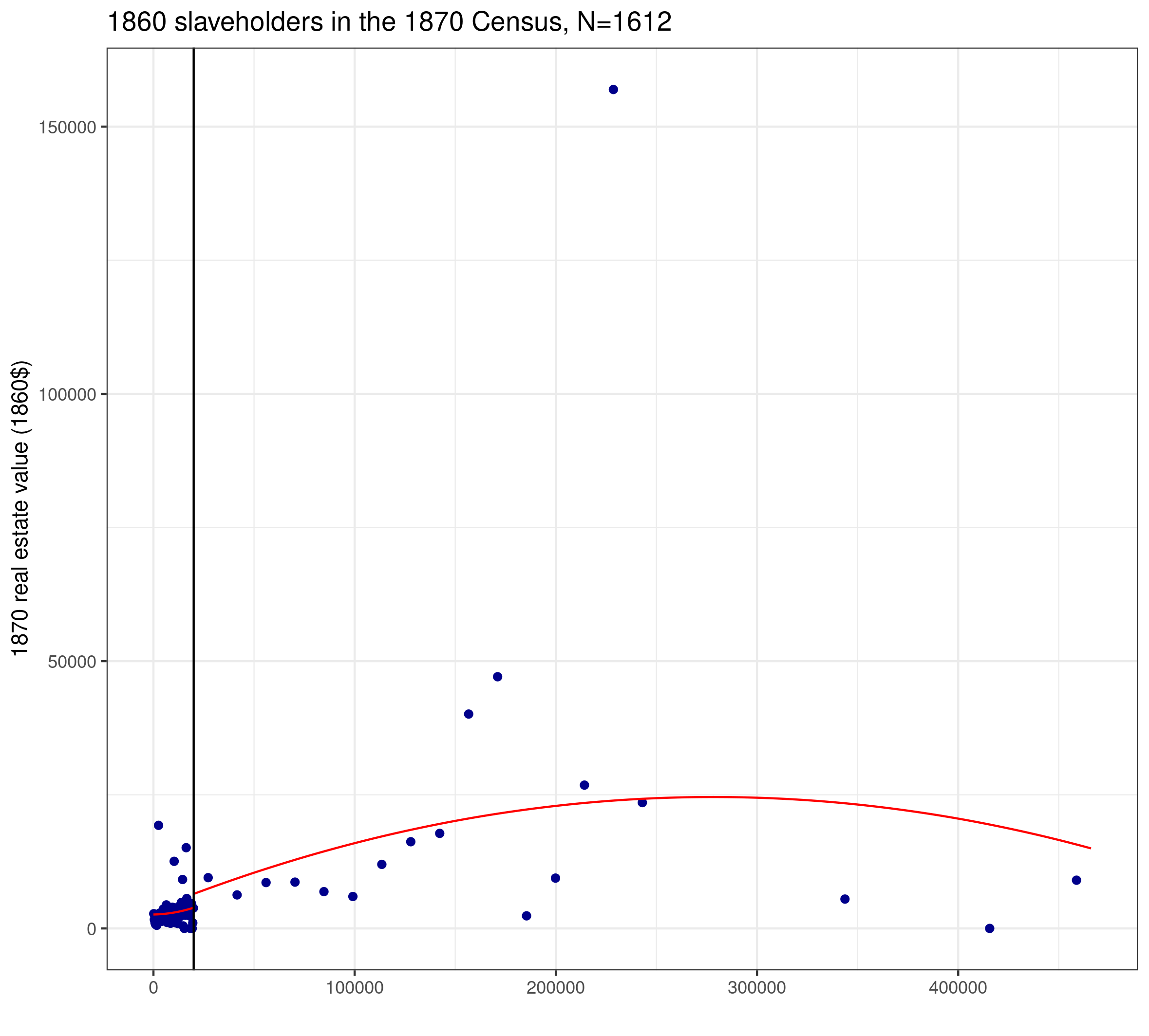} & \\
			\includegraphics[width=.55\textwidth]{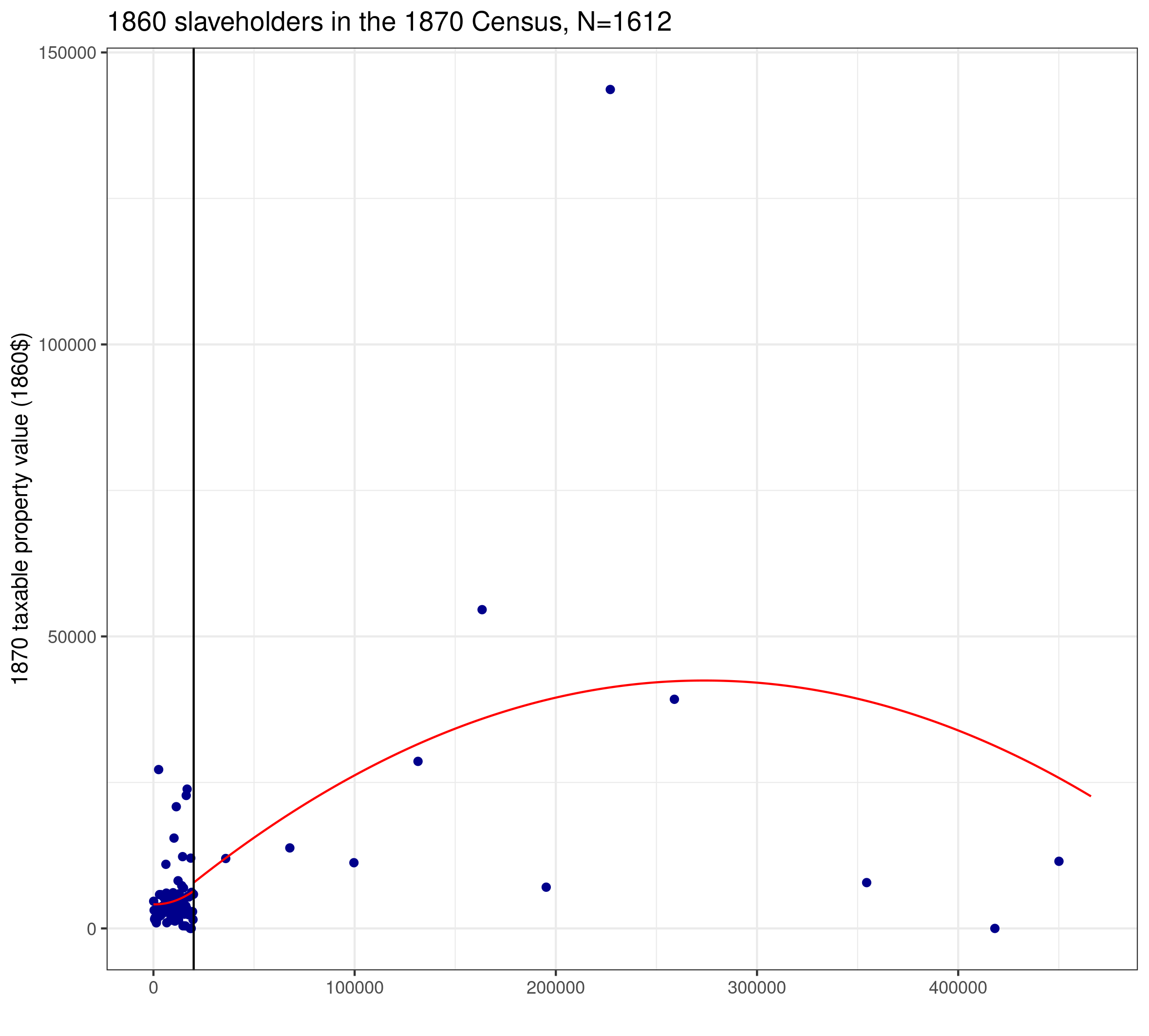} &
		\end{tabular}
		\caption{1860 slaveholders sample: regression discontinuity plots for post-treatment wealth. See notes to Figure \ref{fig:pol-rd}.}
		\label{fig:wealth-rd-slaveholders}
	\end{center}
\end{figure}

\begin{figure}[htbp] 
	\begin{center}
		\includegraphics[width=\textwidth]{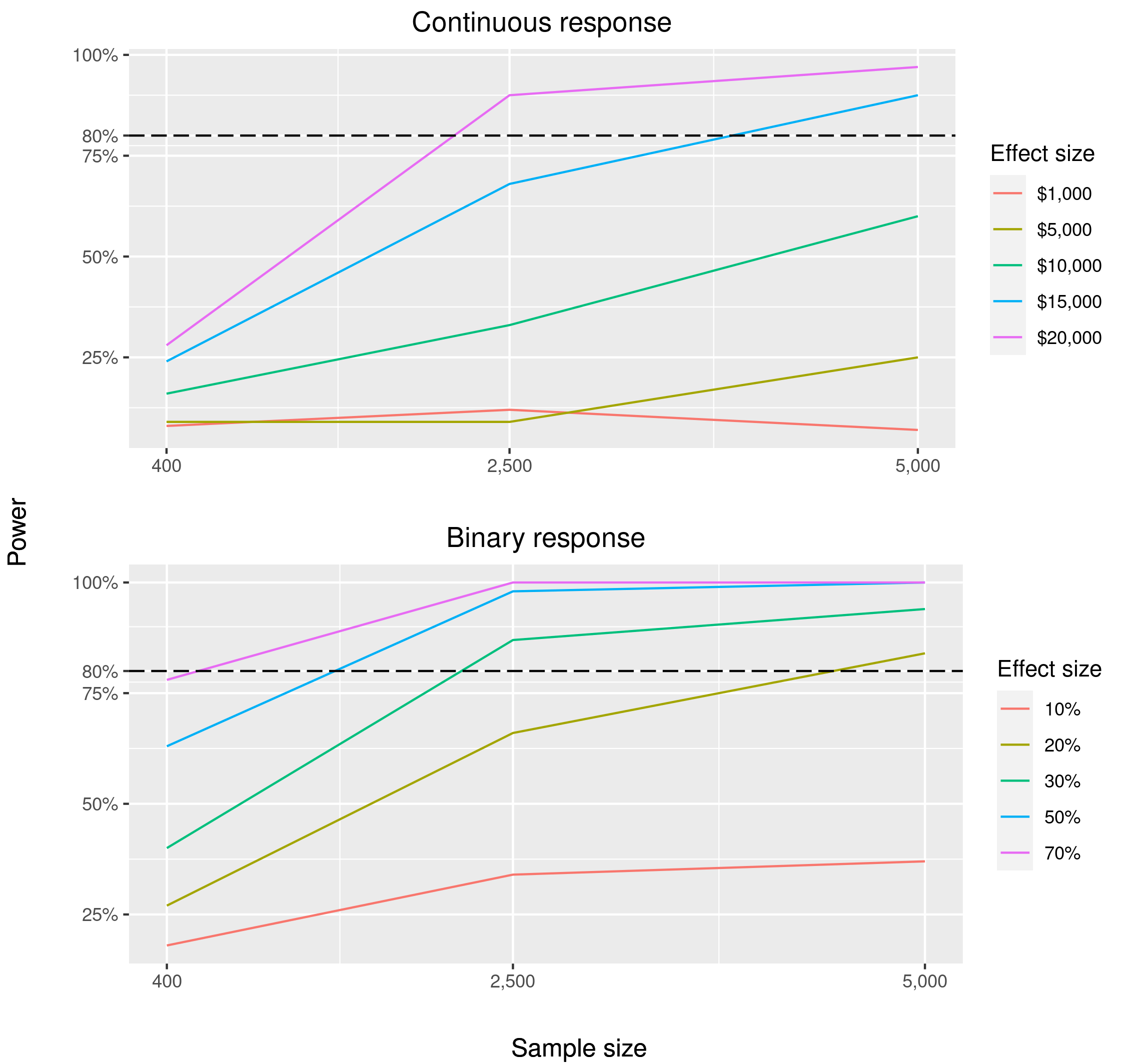} 
		\caption{Power analysis by simulation ($\mathcal{L} =100$ iterations) for estimating Eq. \ref{srd} on continuous and binary variables with varying sample sizes. The horizontal line indicates the power typically needed to justify the study (80\%).}
		\label{fig:power-indiv}
	\end{center}
\end{figure}

\input{rd_estimates_delegates_subgroup.tex}

\newpage
\begin{singlespace}
	\bibliographystyle{chicago}
	\bibliography{references}
\end{singlespace}

\itemize

%% file: data-sources.tex
\begin{table}[]
\small
\centering
\caption{Estimates of the size of the excepted classes.} 
\begin{adjustbox}{max width=\textwidth}
\begin{threeparttable}
\begin{tabular}{@{}lllllll@{}}
\toprule
Dataset/Source                                                      & Sample                                                                                                                                                                                                   & N                                                           & \begin{tabular}[c]{@{}l@{}}\# in 13$^{th}$\\ excep.\end{tabular} & \begin{tabular}[c]{@{}l@{}}\% in 13$^{th}$\\ excep.\end{tabular} & \begin{tabular}[c]{@{}l@{}}\# in all\\ exceps.\end{tabular} & \begin{tabular}[c]{@{}l@{}}\% in all\\ exceps.\end{tabular} \\ \midrule
\begin{tabular}[c]{@{}l@{}}1860 Census \\ (1\% sample)\end{tabular} & \begin{tabular}[c]{@{}l@{}}Adult Srn. males \end{tabular} & \begin{tabular}[c]{@{}l@{}}515,763\end{tabular} & \begin{tabular}[c]{@{}l@{}}67,758\end{tabular}                                 & \begin{tabular}[c]{@{}l@{}} 13\end{tabular}                                & -                                                                             & -                                                                             \\ \\
\rowcolor{Gray}\begin{tabular}[c]{@{}l@{}}1860 Census \\ (5\% slaveholders sample)\end{tabular} & \begin{tabular}[c]{@{}l@{}}Male slaveholders\\ in AL, GA, MS, SC  \end{tabular} & \begin{tabular}[c]{@{}l@{}}5,194\end{tabular} & \begin{tabular}[c]{@{}l@{}}1,329 \end{tabular}                                 & \begin{tabular}[c]{@{}l@{}} 25\end{tabular}                                & -                                                                             & -                                                                             \\ \\
\rowcolor{Gray}\cite{hume2008}                                                   & Reconstruction delegates                                                                                                                                                                                               & 438                                                         & 102                                                                                & 23                                                                             & -                                                                             & -                                                                             \\ \\
\cite{dorris1953}                                                 & Excepted Srns.                                                                                                                                                                                                  & -                                                          & 60-80k                                                                      & $5-6^{*}$                                                       & 150-200k                                                                & $12-16^{*}$                                                    \\ \\
\cite{avery1881}                                                  & Excepted Georgians                                                                                                                                                                               & -                                                          & 12,470                                                                             & $10^{**}$                                                     & 15-20k                                                                  & $12-16^{**}$                                            \\    \bottomrule
\end{tabular}
\begin{tablenotes}
	\item The 13$^{th}$ exception includes individuals with total census wealth exceeding \$20,000.
	\item[] $^*$ denominator is sample of adult southern males in 1860 Census.
	\item[] $^{**}$ denominator is sample of adult males living in Georgia at the time of the 1860 Census.
\end{tablenotes}
\end{threeparttable}
\label{data-sources}
\end{adjustbox}
\end{table}

%% file: rd_estimates_delegates.tex
\begin{table}[]
	  \centering
\begin{adjustbox}{max width=\textwidth}
\begin{tabular}{@{}lll@{}}
\toprule
Outcome                                 & \multicolumn{1}{c}{ITT Estimate}           & \multicolumn{1}{c}{TOT Estimate}                 \\ \midrule
\textit{Binary outcomes}               &                                   &                                   \\
Ex-post officeholder, N=438            & \bf{-0.79 {[}-1.35, -0.30{]}}       & \bf{-0.85 {[}-1.45, -0.32{]}}       \\
Protested constitution, N=173          & -0.05 {[}-0.93, 0.68{]}        & -0.06 {[}-0.99, 0.73{]}        \\
Republican support score, N=314        & -0.42 {[}-2.12, 1.01{]}        & -0.45 {[}-2.27, 1.08{]}        \\
								       &                                   &                                   \\
\textit{1870 Census wealth (1860\$)}        &                                   &                                   \\
Personal property value, N=359 & 1515.52 {[}-5622.90, 7894.46{]}   & 1627.19 {[}-6037.22, 8476.16{]}   \\
Real estate value, N=360       & 10396.70 {[}-2315.38, 21526.54{]} & 11162.78 {[}-2485.99, 23112.71{]} \\
Total census wealth, N=359     & 11574.42 {[}-4100.83, 25804.87{]} & 12427.27 {[}-4402.99, 27706.28{]} \\ \bottomrule
\end{tabular}
\end{adjustbox}
		\caption{Treatment effect estimates for Reconstruction delegates sample. For comparability with the other measures, the Republican support score is normalized to take a value from 0 to 1. Values in brackets represent 95\% confidence intervals are constructed using the standard errors from local polynomial regressions. TOT confidence intervals are obtained by scaling the upper and lower bounds of the ITT confidence interval by $\beta$. Bold values represent statistical significance.}
\label{rd-estimates-delegates}
\end{table}

%% file: rd_estimates_slaveholders.tex
\begin{table}[]
	  \centering
\begin{adjustbox}{max width=\textwidth}
\begin{tabular}{@{}lll@{}}
\toprule
Outcome                                 & \multicolumn{1}{c}{ITT Estimate}           & \multicolumn{1}{c}{TOT Estimate}                 \\ \midrule			\textit{1870 Census wealth (1860\$)} 				       &                                   &                                   \\
Personal property value, N=1,612 & 2966.07 {[}-2218.91, 7812.31{]}   & 3329.32 {[}-2490.65, 8769.06{]}   \\
Real estate value, N=1,612       & 11078.98 {[}-11195.29, 31421.82{]} & 12435.78 {[} -12566.33, 35269.93{]} \\
Total census wealth, N=1,612     & 14096.05 {[}-12669.48, 38942.72{]} &  15822.34 {[}-14221.07, 43711.88{]} \\ \bottomrule
\end{tabular}
\end{adjustbox}
		\caption{Treatment effect estimates for 1860 Slaveholders sample. Values in brackets represent 95\% confidence intervals are constructed using the standard errors from local polynomial regressions.}
\label{rd-estimates-slaveholders}
\end{table}

%% file: chronology.tex
\begin{chronology}[1]{1860}{1870}{15cm}[25cm]
\event[\decimaldate{1}{6}{1860}]{\decimaldate{1}{11}{1860}}{1860 U.S. Census}
\event{\decimaldate{8}{12}{1863}}{PresProc 108 - Amnesty and Reconstruction (Dec. 8, 1863)}
\event[\decimaldate{12}{4}{1861}]{\decimaldate{9}{5}{1865}}{American Civil War (Apr. 12, 1861 - May 9, 1865)}
\event{\decimaldate{29}{5}{1865}}{\textcolor{blue}{{PresProc 134 - Granting Amnesty, with Exceptions (May 29, 1865)}}}
\event{\decimaldate{2}{3}{1867}}{Congress passes first Reconstruction Act (March 2, 1867)}
\event{\decimaldate{1}{9}{1867}}{Election of delegates to Reconstruction conventions (Summer, 1867)}
\event[\decimaldate{5}{11}{1867}]{\decimaldate{6}{2}{1869}}{Reconstruction conventions (Nov. 5, 1867 - Feb. 6, 1869)}
\event{\decimaldate{25}{12}{1868}}{\textcolor{blue}{{PresProc 179 - Granting Full Pardon and Amnesty (Dec. 25, 1868)}}}
\event[\decimaldate{1}{6}{1870}]{\decimaldate{1}{11}{1870}}{1870 U.S. Census} 
\end{chronology}

%% file: slaveholders-sum.tex
{\footnotesize
\begin{longtable}{lrrrrrrr}
 \textbf{Variable} & $\mathbf{N}$ & \textbf{Min} & $\mathbf{Median}$ & $\mathbf{Mean}$ & \textbf{Max} & $\mathbf{SD}$ & \textbf{\#NA} \\ 
  \hline
 &  &      &    &     &     &     &   \\ 
\textit{Wealth measure in 1860 (1860\$):}   &  &      &    &     &     &     &   \\ 
Real estate value & 5,194 &       0 &  1,600 &   5,519 &  381,000 & 14,213 &    0 \\  
Personal property  & 5,194 &       0 &  4,500 &  13,458 &  765,000 & 31,657 &    0 \\ 
Total census wealth & 5,194 &       0 &  6,963 &  18,977 &  781,000 & 41,520 &    0 \\  
&  &      &    &     &     &     &   \\ 
 \textit{Wealth measure in 1870 (1860\$):}   &  &      &    &     &     &     &   \\ 
Real estate value & 1,612 &       0 &   922 &   4,616 &  941,687 & 28,878 & 3,582 \\
Personal property  & 1,612 &       0 &   628 &   2,403 &  313,896 & 13,635 & 3,582 \\ 
Total census wealth & 1,612 &       0 &  1,648 &   7,019 & 1,098,635 & 39,135 & 3,582 \\ 
&  &      &    &     &     &     &   \\ 
 \textit{Biographical:}   &  &      &    &     &     &     &   \\ 
  Age & 5,194 &      21 &    39 &     41.21 &      94 &    13.86 &    0 \\
  &  &      &    &     &     &     &   \\ 
\textit{State of residence:}   &  &      &    &     &     &     &   \\ 
 	Alabama & 5,194 &       0 &     0 &      0.25 &       1 &     0.43 &    0 \\ 
	Georgia & 5,194 &       0 &     0 &      0.29 &       1 &     0.45 &    0 \\ 
	Mississippi & 5,194 &       0 &     0 &      0.23 &       1 &     0.42 &    0 \\ 
	South Carolina & 5,194 &       0 &     0 &      0.24 &       1 &     0.43 &    0 \\ 
	&  &      &    &     &     &     &   \\ 
   \textit{Occupations:}   &  &      &    &     &     &     &   \\ 
	Farmer & 5,194 &       0 &     1 &      0.69 &       1 &     0.46 &    0 \\ 
	Lawyer & 5,194 &       0 &     0 &      0.02 &       1 &     0.15 &    0 \\ 
	Merchant & 5,194 &       0 &     0 &      0.05 &       1 &     0.21 &    0 \\ 
	Physician & 5,194 &       0 &     0 &      0.04 &       1 &     0.20 &    0 \\ 
	Minister & 5,194 &       0 &     0 &      0.01 &       1 &     0.11 &    0 \\ 
	Other & 5,194 &       0 &     0 &      0.19 &       1 &     0.39 &    0 \\
  \hline
\caption{Summary statistics for 1860 slaveholders sample ($N=5,194$).} 
\label{slaveholders-sum}
\end{longtable}
}

%% file: delegates-sum.tex
{\footnotesize
\begin{longtable}{lrrrrrrr}
 \textbf{Variable} & $\mathbf{N}$ & \textbf{Min} & $\mathbf{Median}$ & $\mathbf{Mean}$ & \textbf{Max} & $\mathbf{SD}$ & \textbf{\#NA} \\ 
  \hline
 &  &      &    &     &     &     &   \\ 
\textit{Wealth measure in 1860 (1860\$):}   &  &      &    &     &     &     &   \\ 
Real estate value & 439 &       0 & 2,120 &  6,403 & 194,000 & 15,779 & 135 \\ 
Personal property  &  438&      0 & 3,000 &  9,693 & 170,000 & 18,971 & 136 \\ 
Total census wealth & 438 &       0 & 6,000 & 16,103 & 359,000 & 31,382 & 136 \\ 
&  &      &    &     &     &     &   \\ 
 \textit{Wealth measure in 1870 (1860\$):}   &  &      &    &     &     &     &   \\ 
Real estate value & 455 &       0 & 3,139 &  8,746 & 470,844 & 25,857 & 119 \\ 
Personal property  & 453 &       0 & 1,569 &  5,407 & 470,844 & 27,164 & 121 \\ 
Total census wealth & 453 &       0 & 4,936 & 14,171 & 502,233 & 40,318 & 121 \\ 
&  &      &    &     &     &     &   \\ 
 \textit{Biographical:}   &  &      &    &     &     &     &   \\ 
  Age & 543 &       23 &   45 &    45.70 &     82 &    11.24 &  31 \\ 
  Confederate & 574 &       0 &    0 &     0.06 &      1 &     0.24 &   0 \\ 
    Democrat  & 574 &    0 &    0 &     0.04 &      1 &     0.20 &   0 \\ 
Former officeholder& 574 &     0 &    0 &     0.23 &      1 &     0.42 &   0 \\ 
  Unionist & 574 &      0 &    0 &     0.12 &      1 &     0.32 &   0 \\ 
  &  &      &    &     &     &     &   \\ 
   \textit{District characteristics:}   &  &      &    &     &     &     &   \\ 
  Percent black & 574 &      0 &   44.50 &    42.70 &     89.40 &    20.56 &   0 \\ 
  &  &      &    &     &     &     &   \\ 
   \textit{Occupations:}   &  &      &    &     &     &     &   \\ 
  Farmer & 574 &        0 &    0 &     0.38 &      1 &     0.49 &   0 \\ 
  Lawyer & 574 &       0 &    0 &     0.21 &      1 &     0.41 &   0 \\ 
    Merchant & 574 &        0 &    0 &     0.08 &      1 &     0.27 &   0 \\ 
  Physician & 574 &        0 &    0 &     0.11 &      1 &     0.32 &   0 \\ 
  &  &      &    &     &     &     &   \\ 
 \textit{Political response variables}   &  &      &    &     &     &     &   \\ 
  Ex-post officeholder & 574 &         0 &    0 &     0.49 &      1 &     0.50 &   0 \\
  Protested constitution & 234 &        0 &    0 &     0.31 &      1 &     0.46 & 340 \\
  Republican support score& 398 &       0.00 &    2.59 &     2.50 &      5.00 &     1.70 & 176 \\  
  \hline
\caption{Summary statistics for sample of Reconstruction delegates ($N=574$).} 
\label{delegates-sum}
\end{longtable}
}

%% file: rd_estimates_delegates_subgroup.tex
\begin{table}[]
	  \centering
\begin{adjustbox}{max width=\textwidth}
\begin{tabular}{@{}lll@{}}
\toprule
Outcome                                 & \multicolumn{1}{c}{Ex-post officeholder}           & \multicolumn{1}{c}{Ex-post officeholder}                 \\ 
                                & \multicolumn{1}{c}{(ITT)}           & \multicolumn{1}{c}{(TOT)}                 \\ \midrule
\textit{Pretreatment covariates}       &                                   &                                   \\
Confederate, N=24            & 0.39 {[}-0.99, 1.78{]}       & 0.45 {[}-1.13, 2.04{]}       \\
Democrat, N=18          & 0.51 {[}-1.13, 2.16{]}        & 0.64 {[}-1.41, 2.70{]}        \\
Former officeholder, N=110        &  -0.53 {[}-2.15,  1.07{]}        & -0.53 {[}-2.15,  1.07{]}         \\
Unionist, N=47        &  -0.06 {[}-1.39, 1.26{]}        & -0.06 {[}-1.50,  1.36{]}        \\
\bottomrule
\end{tabular}
\end{adjustbox}
		\caption{Subgroup treatment effect estimates on ex-post officeholding for Reconstruction delegates sample. See notes to Table \ref{rd-estimates-delegates}.}
\label{rd-estimates-delegates-subgroup}
\end{table}